\documentclass[5p,twocolumn]{elsarticle}

\usepackage{amssymb,amsthm}
\usepackage{amsmath}

\usepackage[dvips]{color}
\usepackage{graphicx} 

\usepackage{pdfsync}

\usepackage{tikz}
\usepackage{xcolor}
\usetikzlibrary{arrows}

\usepackage{caption}

\journal{Journal of Theoretical Biology}

\begin{document}

\begin{frontmatter}



\title{Optimal allocation patterns and optimal seed mass of a perennial plant}

\author[wue]{Andrii Mironchenko\corref{cor1}}
\ead{andrii.mironchenko@mathematik.uni-wuerzburg.de}

\author[jag]{Jan Koz\l owski}
\ead{jan.kozlowski@uj.edu.pl}

\cortext[cor1]{Corresponding author}

\address[wue]{Institute of Mathematics, University of W\"urzburg, Emil-Fischer Stra\ss e 40,
97074 W\"urzburg, Germany}
\address[jag]{Institute of Environmental Sciences, Jagiellonian University, Gronostajowa 7, 30-387 
Krak\'ow, Poland}


\begin{abstract}
We present a novel optimal allocation model for perennial plants, in which assimilates are not allocated directly to vegetative or reproductive parts but instead go first to a storage compartment from where they are then optimally redistributed. We do not restrict considerations purely to periods favourable for photosynthesis, as it was done in published models of perennial species, but analyse the whole life period of a perennial plant. As a result, we obtain the general scheme of perennial plant development, for which annual and monocarpic strategies are special cases.

We not only re-derive predictions from several previous optimal allocation models, but also obtain more information about plants' strategies during transitions between favourable and unfavourable seasons. One of the model's predictions is that a plant can begin to re-establish vegetative tissues from storage, some time before the beginning of favourable conditions, which in turn allows for better production potential when conditions become better.
By means of numerical examples we show that annual plants with single or multiple reproduction periods, monocarps, evergreen perennials and polycarpic perennials can be studied successfully with the help of our unified model.

Finally, we build a bridge between optimal allocation models and models describing trade-offs between size and the number of seeds: a modelled plant can control the distribution of not only allocated carbohydrates but also seed size. We provide sufficient conditions for the optimality of producing the smallest and largest seeds possible.

\end{abstract}


\end{frontmatter}

\section{Introduction}

The pioneering work \cite{Coh1971} gave rise to a new class of mathematical models of plants based on methods of optimal control theory. In these models it was assumed that a plant can control resource allocation in order to maximise its fitness, which is often identified with the mass of seeds produced by a plant during its lifetime.

In the first models, which were devoted to the development of annual plants, it was assumed that a plant consists of a number of compartments -- at least of a vegetative compartment (leaves, roots, stems) and a reproductive compartment (seeds and auxiliary tissues), although storage and defensive tissues could also be included. 

This basic model, posited by \cite{ViP1980}, results in a bang-bang transition from the allocation to vegetative tissues to the allocation to seeds. This annual plant model has been extended in many directions, in particular
in \cite{IwR1984} a model with multiple vegetative compartments was analysed, and in \cite{ZiK1995} and \cite{IoG2005}, in which additional physiological constraints were considered, resulting in periods of mixed growth (where both the vegetative and reproductive parts of a plant grow simultaneously). 
Optimal allocation strategies in stochastic environments have been investigated in particular in \cite{Lar2006}, while allocation to defensive tissues was encountered in \cite{YFH2007} and \cite{TaY2010}, to cite a few examples.

The survey of early works in this field is provided in \cite{Fox1992}, and for a general overview of resource allocation in plants see books \cite{BaG1997} and \cite{ReB2005}.

In contrast to annual plants, less attention has been devoted to the modelling of perennials' optimal phenology. Usually, the behaviour of a perennial plant is modelled in the following way: its lifetime is divided into discrete seasons during which environmental conditions are favourable for photosynthesis. 
The model of a plant within every season is continuous and is treated with the methods used in annual plant models \cite{Sch1983}, \cite{IwC1989}, \cite{PuK1990}.
To model the behaviour of a plant between seasons (when the weather is unfavourable), some simple transition rules are used that show which parts of compartments are saved during the season and which are not. The solution to such problems is divided into two parts: first, the model on one season is solved using Pontryagin's Maximum Principle (see e.g. \cite{ATF87}), and then a solution to the whole model is sought by applying the dynamic programming method.

Although these models provide quite interesting qualitative results regarding the behaviour of perennial plants, they have an important disadvantage, namely that the subtle qualitative behaviour of a plant during a season contrasts with the simple jump from the end of one season to the beginning of the next one. In this paper we propose a perennial plant continuous-time model, which allows us to describe more precisely the dynamics of a plant during seasons with unfavourable environmental conditions for photosynthesis, and to avoid the introduction of additional parameters for describing jumps between seasons. 

With the help of Pontryagin's Maximum Principle we derive a general scheme of perennial plant development, which contains models for annual as well as monocarpic plants as special cases. We also prove that monocarpy is always optimal if there are no losses of storage parts and there is no mortality before the end of life.

With the help of numerical examples we show that many developmental patterns from previous papers can be derived by using our model. In particular, one can use it to study annual plants with multiple reproduction periods \cite{KiR1982}; perennial plants which grow to a certain size for a number of periods, and in the following periods they firstly regrow to this size and then produce reproductive tissues \cite{IwC1989}; the evergreen polycarpic plants as well as monocarps.
Moreover, our model is much better suited for the study of transitions from favourable to unfavourable climate conditions, and one of its predictions is that plants begin to generate vegetative tissues not at a time when environmental conditions are favourable for photosynthesis, but slightly earlier in order to enter into the suitable period with developed vegetative tissues.

Having established an optimal allocation model we will connect it to the theory on trade-offs between size and number of seeds.
A lot of attention has been devoted to these trade-offs in the scientific literature.
The basic model has been proposed in a seminal work \cite{SmF1974}, where it was assumed that the fitness of a plant is equal to the sum of the fitnesses of its descendants.
Afterwards this model has been generalised in a number of directions (for a review see \cite{Fen2000}). 
In this framework, optimal size is sought depending on the properties of the fitness function. This makes possible the quite general treatment of size-number trade-offs, but the question remains as to how to formalise the dependency of fitness on size and the number of seeds and how to find the properties of the function that characterises this dependency.

Our aim is to investigate the trade-offs between the number and size of seeds in the context of
optimal allocation models. Within this framework fitness is properly formalised, and we can 
investigate the optimal size of a seed depending on the properties of the photosynthetic rate function and other physiological parameters of a plant, which offer more distinct criteria than abstract fitness. We provide the analysis for the model developed in Section \ref{HauptModell} of this paper, but the results are valid also for a number of other optimal allocation models. We prove that, according to our plant model, if the photosynthetic rate function is concave (that is, if the rate of photosynthesis per unit mass decays with an increase in the size of a plant), then the seeds have to be as small as possible. Such behaviour is particularly typical in colonising species (see Section \ref{Besprechung}).

Our model also includes the possibility of choosing the germination time of a seed. 
As a consequence, we obtain results concerning the behaviour of plants from dormancy of seeds up to senile stage.

The outline of the article is as follows. In Section \ref{S:Description} we introduce a perennial plant model. In Section~\ref{S:Results} we summarise the predictions of the model, provide a general plant development scheme and consider a number of special cases (annual and monocarpic plants). In Section~\ref{sec:Numerics} we make numeric simulations of different plant development scenarios, while in Section \ref{Section_SamenMass} we consider trade-offs between size and the number of seeds. 
The results of the paper are discussed in Section \ref{Besprechung}, and Section \ref{Schluss} draws conclusions from the results and outlines some directions for future work. In Appendices~1 and 2 we provide derivations of our theoretical results.

\section{Optimal allocation model}
\label{HauptModell}

\subsection{Model description}
\label{S:Description}

In optimal allocation models it is usually assumed that all allocated photosynthate is immediately used for the construction of tissues. In models taking into account the presence of a storage compartment, a plant can also allocate resources from storage, depending on the mass of the storage. Such a method ignores the fact that a photosynthate is not immediately allocated to certain structures but instead exists for some time in a free state. We shall take this effect into account and assume that an intermediate stage exists whereby carbohydrates have already been photosynthesised but have not yet been permanently allocated to a given structure.

Let a plant consist of three parts: a vegetative compartment, a reproductive compartment and non-structural carbohydrates (free glucose, starch, etc.), hereinafter called `storage'. Let
\begin{itemize}
\item $x_1(t)$ be the mass of the vegetative compartment at time $t$, 
\item  $x_2(t)$ be the mass of the reproductive compartment at time $t$, 
\item  $x_3(t)$ be the mass of storage at time $t$.
\end{itemize}

We model the dynamics of a plant via the following equations:
\begin {equation}
\label{FreiFotModell}
\begin {array} {l}
\dot{x}_1 = v_1(t) g(x_3) - \mu(t)x_1, \\
\dot{x}_2 = (v(t)-v_1(t)) g(x_3),   \\
\dot{x}_3 = \zeta(t)f(x_1) - v(t) g(x_3) - \omega(t) x_3.
\end {array} 
\end {equation}

Here $f(x_1)$ describes the rate of photosynthesis of the plant with vegetative mass $x_1$ in optimal environmental conditions, and $g(x_3)$ - the maximal rate of allocation of non-structural carbohydrates. It is natural to assume that $f$ and $g$ increase monotonically and $f(0)=g(0)=0$. 

Climate influence is modelled by three functions: 
$\zeta: [0,T] \to [0,1]$ and $\mu, \omega: [0,T] \to [0,\infty)$, where $T$ is maximum longevity. 
\begin{itemize}
\item $\zeta(t)$ models the dependence of the rate of photosynthesis on the climate ($\zeta(t) = 0$ if at time $t$ no photosynthesis is possible); 
\item $\mu(t)$ is the loss rate of vegetative tissues per unit mass at time $t$;
\item $\omega(t)$ is the loss rate of the storage parts per unit mass due to external factors (decaying, grazing by animals, etc.) at time $t$.
\end{itemize}

Note that photosynthesised carbohydrates firstly enlarge the mass of storage before they can be allocated to other compartments.

We assume that a plant can control the total allocation rate with the control $v(t) \in [0,1]$ and the allocation rate to vegetative tissues with the control $v_1(t) \in [0,v(t)]$; consequently, the allocation rate to reproductive tissues at time $t$ is controlled by $v_2(t)=v(t)-v_1(t)$, and $v(t)=0$ means that resources are not being relocated from storage at time $t$.

The initial mass of the seed and all its compartments is given as follows
\begin {equation}
\label{AnfangsBedingungen}
x_i(0)=x_i^0, \quad i=1,2,3.
\end {equation}
Problem of optimisation of a seed mass will be considered in Section~\ref{Section_SamenMass}.

Seed dormancy is modelled as the ability of a plant to choose the time of germination $t_0 \in [0,T]$. For simplicity we assume that a seed cannot decay and it does not use any resources for life-sustaining activities before germination. Thus
\begin {equation}
\label{SproutingBedingungen}
x_i(t_0)=x_i(0)=x_i^0, \quad i=1,2,3.
\end {equation}

To model the mortality of a parental plant, we introduce the function $\tilde L:[0,T] \to [0,1]$. $\tilde L(t)$ is the probability of survival of a parental plant to age $t$. We assume in this paper that mortality is only age-dependent and does not depend on the size of a plant. Since the time of germination may vary, it makes sense to introduce the function $L_{t_0}$, defined by relation $L_{t_0}(t)=\tilde L(t_0-t)$. In what follows, we write for short $L=L_{t_0}$.

It is natural to assume that $L$ is a non-increasing function and that $L(t)>0$ for all $t \in [0,T)$. In fact, if $L(t) \equiv 0$
on $[T-\varepsilon,T]$ for some $\varepsilon>0$, then this means that at moment $T-\varepsilon$ a plant will already be dead, and we can therefore consider the optimal control problem on time-period $[0,T-\varepsilon]$.

We choose the maximisation of the expected total yield of seeds over a lifespan as the fitness measure. Thus: 
\begin {equation}
\label{Ziel}
\int_{t_0}^T L(s) \dot{x}_2(s) ds = \int_{t_0}^T L(s) (v(s)-v_1(s)) g(x_3(s)) ds  \rightarrow \max.
\end {equation}

A plant can maximise fitness by choosing an appropriate germination time, $t_0$, and controls $v$ and $v_1$ defined on $[t_0,T]$.

We assume that the functions on the right-hand side of equations \eqref{FreiFotModell} are smooth enough to guarantee the existence and uniqueness of solutions for \eqref{FreiFotModell}. We also assume that the system \eqref{FreiFotModell} is forward complete, i.e. for all initial conditions and all admissible controls, the solution of \eqref{FreiFotModell} exists for all time. From the biological viewpoint this means that it is impossible to achieve endless yields over a finite amount of time, which in turn ensures that the solution to the problem \eqref{FreiFotModell}, \eqref{AnfangsBedingungen}, \eqref{Ziel} exists.

\subsection{Model predictions}
\label{S:Results}

We perform an analysis of \eqref{FreiFotModell}, \eqref{Ziel} with the help of 
Pontryagin's Maximum Principle (PMP) \cite{ATF87}, with the aim of determining in what order the periods of a plant's life follow each other. 
The full analysis is presented in Appendix~1, but for now we summarise the results derived therein.

First, we construct a Hamiltonian $H$, corresponding to the problem \eqref{FreiFotModell}, \eqref{Ziel}, which is equal to the scalar product of the right-hand side of \eqref{FreiFotModell} times the functional coefficients $p_1, L, p_3$.
\begin {equation*}
\begin{array}{l}
{H= p_1(t) \left(v_1(t) g(x_3(t)) - \mu(t)x_1(t)\right)} \\
   {\phantom{aaaaa} + L(t)\left(v(t)-v_1(t)\right) g(x_3(t))}\\
{\phantom{aaaaaa}  +p_3(t) \left(\zeta(t)f(x_1(t)) {-} v(t) g(x_3(t)) {-} \omega(t) x_3(t)\right).}
\end{array}
\end {equation*}
The coefficients $p_1$ and $p_3$ are so-called `adjoint functions' to the equations, as they govern the dynamics of $x_1$ and $x_3$. Their dynamics is described by equations \eqref{P-Gleichungen}. 

The development of a plant according to the model \eqref{FreiFotModell} involves three main periods: 
\begin{enumerate}
	\item[(V)] Vegetative period, during which holds
\[
p_1(t)>\max\{ L(t),p_3(t)\}. 
\]
	\item[(R)] Reproductive period, characterised by 
\[
 L(t)>\max\{p_1(t),p_3(t)\}.
\]
	\item[(S)] Storage period, during which 
\[
p_3(t)>\max\{p_1(t), L(t)\}. 
\]
\end{enumerate}

Thus, the choice of a compartment to which a plant should allocate available resources depends on the relation between functions $p_1, L, p_3$: all resources should be allocated to vegetative parts if $p_1$ is the largest of the trio, to storage if $p_3$ is the largest and to reproduction if $L$ is the largest.

The periods $V$, $R$, $S$ can be further subdivided into sub-periods which follow each other, as depicted in Figure \ref{fig:DiagPlantsBIG}:
\begin{itemize}
	\item $D$ - Seed dormancy.
	\item $S.2$ - Preparing for unfavourable climate conditions.
	\item $S.1$ - Life in unfavourable climate conditions.
	\item $V.2$ - Vegetative period that starts close to the end of the period with unfavourable conditions. At the beginning of this period, a plant starts allocating to vegetative tissues as preparation for climate conditions which are favourable for photosynthesis.
	\item $V.1$ Allocation to vegetative tissues before reproduction.
\end{itemize}

\begin{figure*} 
	\centering
\caption*{A: perennial plants}
\begin{tikzpicture}[text width=1cm, text centered]

\node [rectangle, draw](A1)   at (0,0) {$D$};
\node [rectangle, draw](A2)   at (2,0) {$V$};
\node [rectangle, draw](A3)   at (4,0) {$S.2$};
\node [rectangle, draw](A4)   at (6,0) {$S.1$};
\node [rectangle, draw](A56)   at (8,0) {$V.2$};
\node [rectangle, draw](A7)   at (10,0) {$V.1$};
\node [rectangle, draw](A8)   at (12,0) {$R$};

\draw[->] (A1) to (A2);
\draw[->] (A2) to (A3);
\draw[->] (A3) to (A4);
\draw[->] (A4) to (A56);

\draw[->] (A56) to (A7);
\draw[->] (A7) to (A8);

\draw[->] (A2) to (2,0.7) to (12,0.7) to (A8);

\draw[->] (A8) to (12,-0.7) to (4,-0.7) to (A3);
\draw[->] (A8) to (12,-0.5) to (10,-0.5) to (A7);

\draw[->] (7.7,0.25) to (7.7,0.5) to (4,0.5) to (A3);

\draw[->] (A7) to (10,0.5) to (8.3,0.5) to (8.3,0.25);

\end{tikzpicture}
\caption*{B: annual plants}
\begin{tikzpicture}[text width=1cm, text centered]

\node [rectangle, draw](A1)   at (0,0) {$D$};
\node [rectangle, draw](A2)   at (2,0) {$V$};
\node [rectangle, draw](A3)   at (4,0) {$V.1$};
\node [rectangle, draw](A8)   at (6,0) {$R$};

\draw[->] (A1) to (A2);
\draw[->] (A3) to (A8);
\draw[->] (A8) to (6,-0.7) to (4,-0.7) to (A3);
\draw[->] (A2) to (2,0.7) to (6,0.7) to (A8);
\end{tikzpicture}
\caption*{C: monocarpic plants}
\begin{tikzpicture}[text width=1cm, text centered]

\node [rectangle, draw](A1)   at (0,0) {$D$};
\node [rectangle, draw](A2)   at (2,0) {$V$};
\node [rectangle, draw](A3)   at (4,0) {$S.2$};
\node [rectangle, draw](A4)   at (6,0) {$S.1$};
\node [rectangle, draw](A56)   at (8,0) {$V.2$};
\node [rectangle, draw](A7)   at (10,0) {$V.1$};
\node [rectangle, draw](A8)   at (12,0) {$R$};

\draw[->] (A1) to (A2);
\draw[->] (A2) to (A3);
\draw[->] (A3) to (A4);
\draw[->] (A4) to (A56);

\draw[->] (A56) to (A7);
\draw[->] (A7) to (A8);

\draw[->] (A2) to (2,0.7) to (12,0.7) to (A8);

\draw[->] (A8) to (12,-0.5) to (10,-0.5) to (A7);

\draw[->] (7.7,0.25) to (7.7,0.5) to (4,0.5) to (A3);
\end{tikzpicture}
\vspace{5mm}

\caption{Stages of development of a perennial (A), annual (B) and monocarpic (C) plant. \textit{D} stays for seed dormancy, \textit{V} -- vegetative growth after germination, \textit{S}.2 -- Preparing for unfavourable climate conditions, \textit{S}.1 -- life in unfavourable climatic conditions, \textit{V}.2 -- vegetative period, \textit{V}.1 -- allocation to vegetative tissues before reproduction, \textit{R} -- reproductive allocation.} 
	\label{fig:DiagPlantsBIG}
\end{figure*}

Important special cases of the general scheme are:
\begin{enumerate}
	\item Annual plant with the possibility of multiple reproduction periods (Figure~\ref{fig:DiagPlantsBIG} B). Multiple reproduction periods may appear if losses of vegetative mass due to external factors modelled by the function $\mu$ are severe. This particular case was analysed in the early work \cite{KiR1982}. We show a numerical example of this scenario in Section \ref{Numerics_Annuals}. If $\mu \equiv 0$, then multiple reproductive periods for annual plants are not possible. 

	\item Monocarpic plants. A sufficient (but not necessary) condition for a plant to be monocarpic (if there is no mortality) is the negligibility of $\omega$ (in particular, if $\omega \equiv 0$); in other words, the mass of storage cannot decrease due to external factors. In this case, transitions $R$ $\to$ $S.2$ and $V.1$ $\rightarrow$ $V.2$ are not possible, as shown in Figure \ref{fig:DiagPlantsBIG} C. We show a numerical example of this scenario in Section \ref{Numerics_Monocarps}.
\end{enumerate}

\section{Numerical examples}
\label{sec:Numerics}

In this section we present examples of numerical solutions for the model \eqref{FreiFotModell}, which represents various patterns of plant life histories. The examples are not intended to mimic any specific plant species but to show that the model is adequate enough to include a broad range of strategies previously modelled separately. Additionally, we show that parameter changes may lead to qualitatively different results.  

We start with annual plants with single or multiple periods of reproductive allocation, following which we present the cases of monocarpic and evergreen polycarpic plants as well as a polycarpic plant losing almost all of its vegetative parts
but retaining its storage during unfavourable seasons. Finally, we show that annual or monocarpic strategies can evolve under certain conditions, even when lifespan is not predefined.

In all examples time is measured in months, with the beginning and the end of the simulation placed in the middle of a winter. The mass of all three compartments (vegetative mass, storage, reproductive mass) is measured in energy units, say MJ, to avoid using coefficients, in order to take into account water content and differences in energy density between compartments. Seeds of the size 0.3 units contain 95 per cent of storage and 5 per cent of vegetative mass. The photosynthetic rate is described via the following saturation function:
\begin{equation}
\label{PhotoRate_Form}
f(x) = \frac{ax}{bx+k},
\end{equation}
where $a,b,k > 0$ (\cite[p. 224]{ViP1980}). Another reasonable choice could be an allometric function $ax^b$, used in particular within Metabolic Theory of Ecology (MTE), Dynamic Energy Budgets (DEBs), etc. (see \cite{Mee06}, \cite{Lav82}).

We model the dependence of the photosynthetic rate on climate as follows:
\[
\zeta(t):= 0.2 + 0.8 \left|\sin\left(\frac{\pi}{12}t\right)\right|.
\]

The maximal release rate of storage tissues is linear, i.e. $g(x_3)=c x_3$ for some $c>0$ -- while the actual release rate depends on the control variable $v(t)$. Storage and vegetative mass losses are defined separately for particular cases. 

If the contrary is not mentioned explicitly, we assume that there is no mortality ($L \equiv 1$).

Computations are made in Matlab, with the help of the optimal control solver GPOPS.

\subsection{Annual plants}
\label{Numerics_Annuals}

\begin{figure*}
\begin{minipage}{0.5\linewidth}
\centering
\caption*{A: Annual with one reproduction period: states} 
\vspace{-3mm}
\includegraphics[scale=0.45]{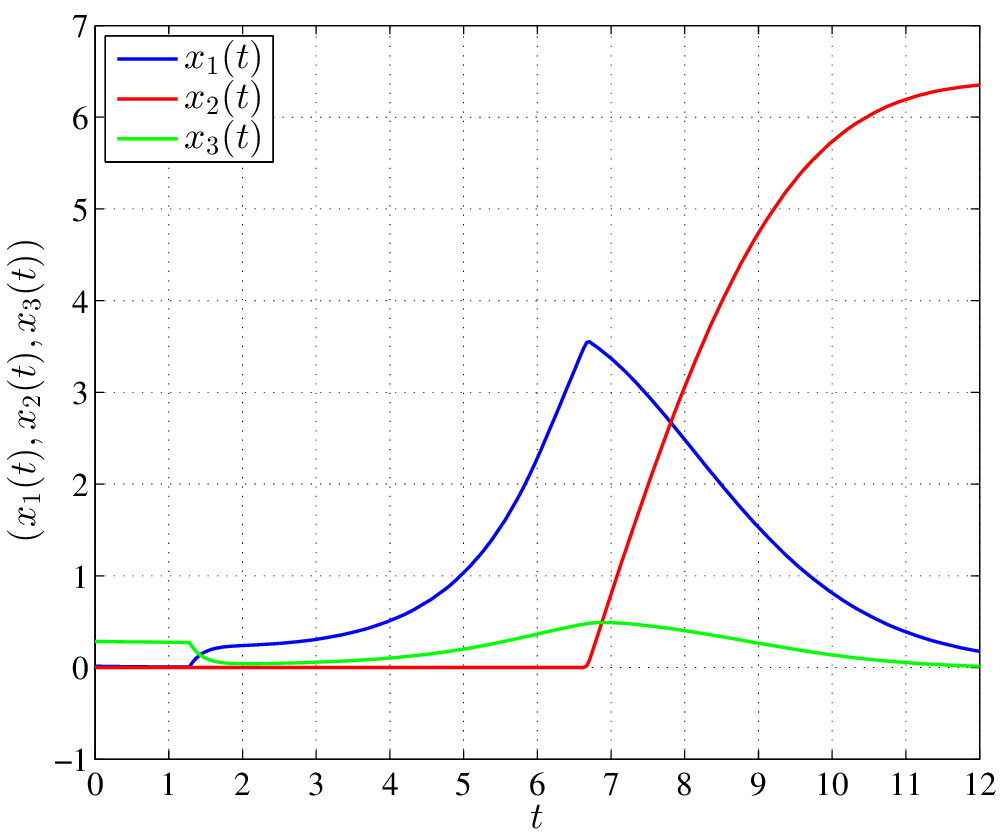}
\label{fig:MirKozState_Annual_Single}
\end{minipage}
\begin{minipage}{0.5\linewidth}
\centering
\caption*{B: Annual with multiple reproduction periods: states}    
\vspace{-3mm}
\includegraphics[scale=0.45]{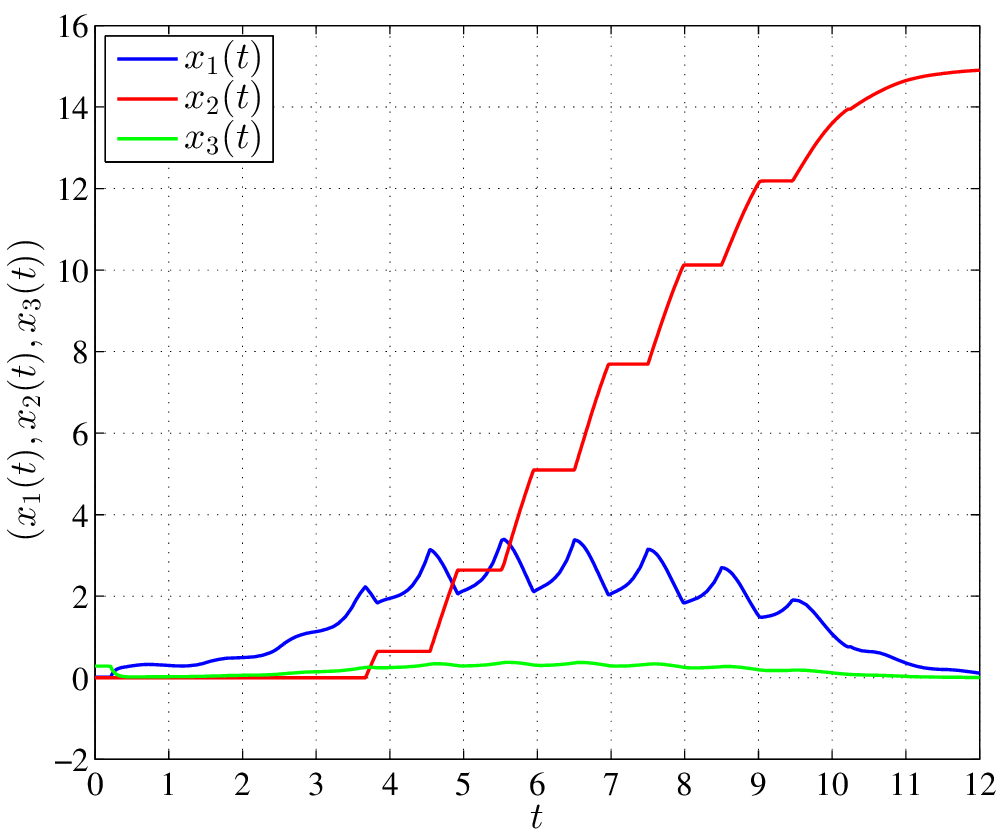}
\label{fig:MirKozState_Annual_Multiple}
\end{minipage}

\begin{minipage}{0.5\linewidth}
\setlength\parindent{0pt}
\centering
\vspace{5mm}
\caption*{C: Annual with one reproduction period: controls}    
\vspace{-3mm}
\includegraphics[scale=0.45]{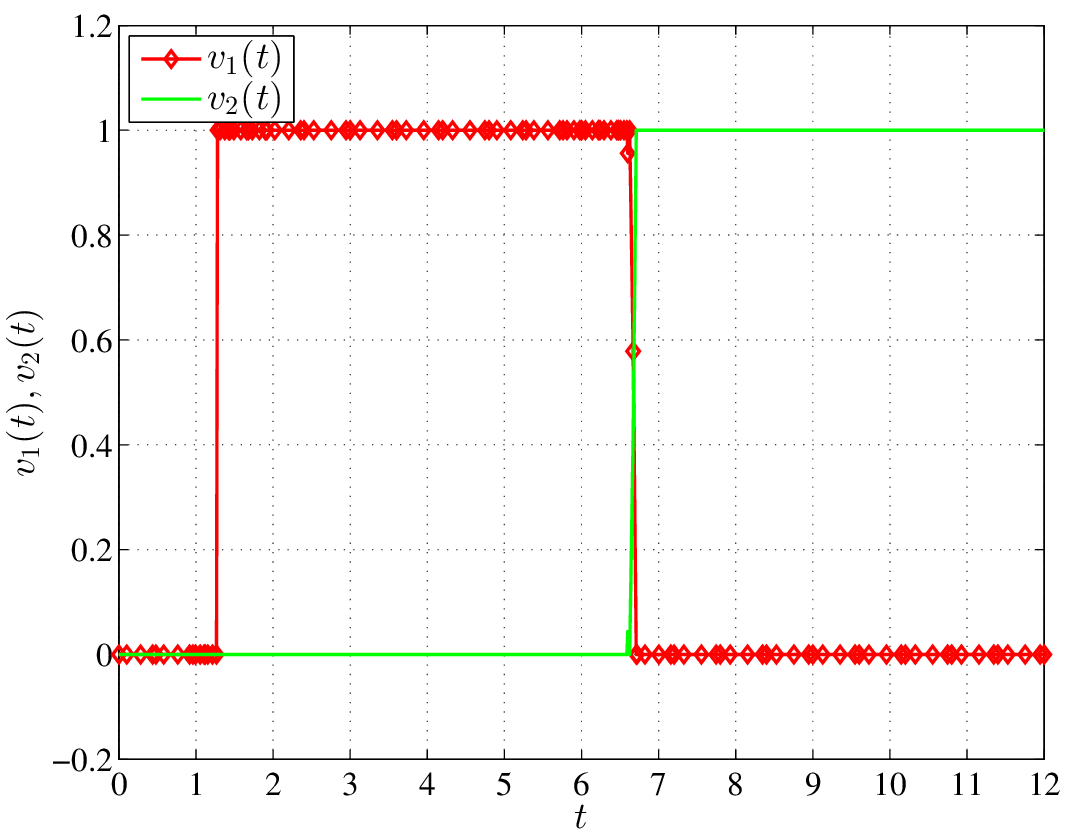}
    \label{fig:MirKozControl_Annual_Single}
\end{minipage}
\begin{minipage}{0.5\linewidth}
\setlength\parindent{0pt}
\centering
\vspace{5mm}
\caption*{D: Annual with multiple reproduction periods: controls}    
\vspace{-3mm}
\includegraphics[scale=0.45]{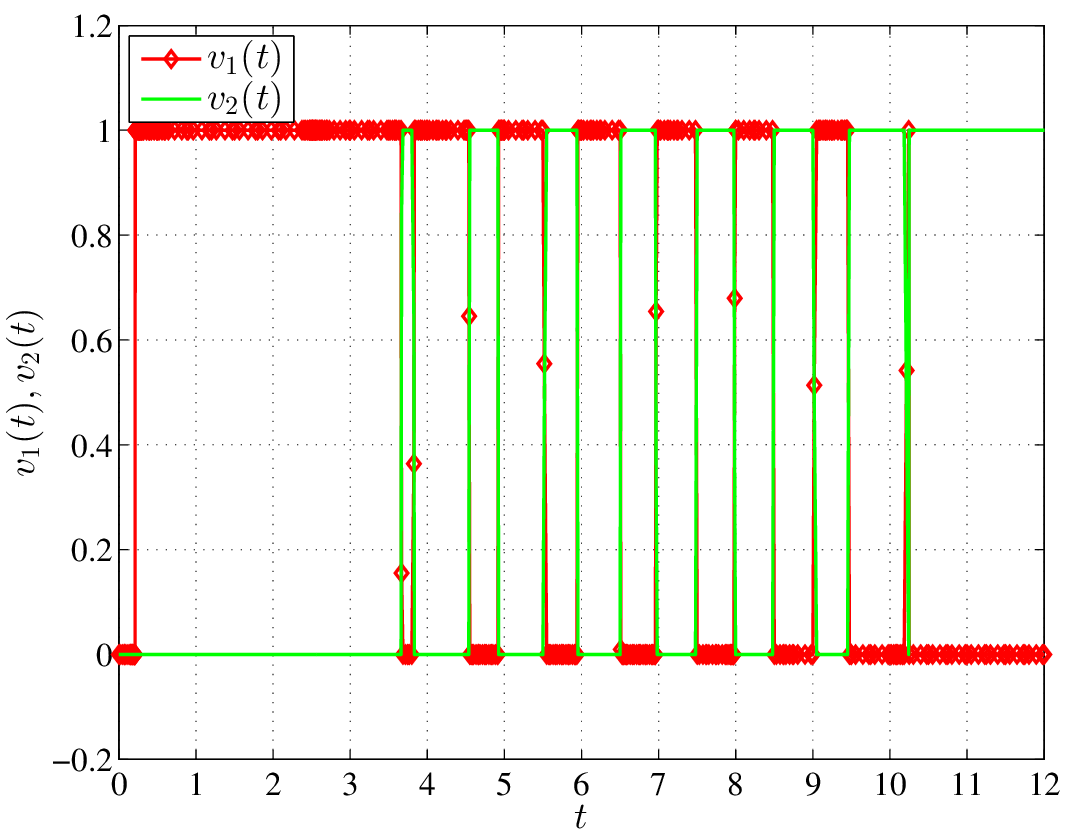}
    \label{fig:MirKozControl_Annual_Multiple}    
\end{minipage}

\begin{minipage}{0.5\linewidth}
\centering
\vspace{5mm}
\caption*{E: Annual with one reproduction period: costates}    
\vspace{-3mm}
\includegraphics[scale=0.45]{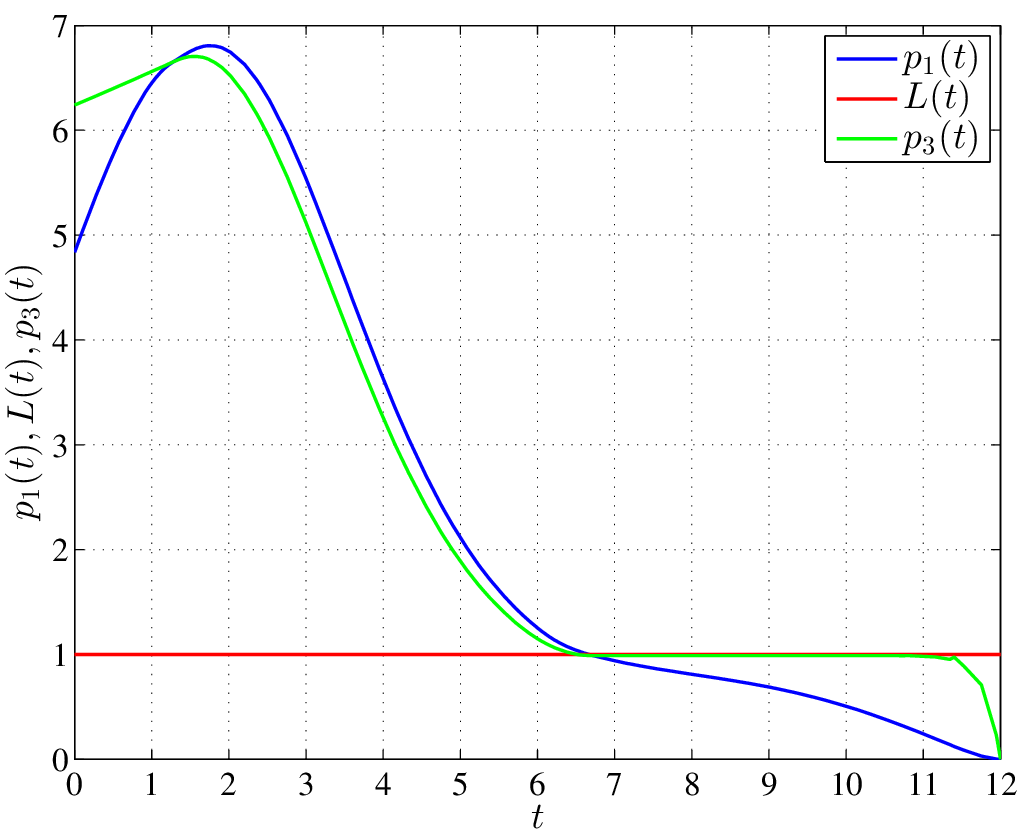}
\label{fig:MirKozCostate_Annual_Single}
\end{minipage}
\begin{minipage}{0.5\linewidth}
\setlength\parindent{0pt}
\centering
\vspace{5mm}
\caption*{F: Annual with multiple reproduction periods: costates}    
\vspace{-3mm}
\includegraphics[scale=0.45]{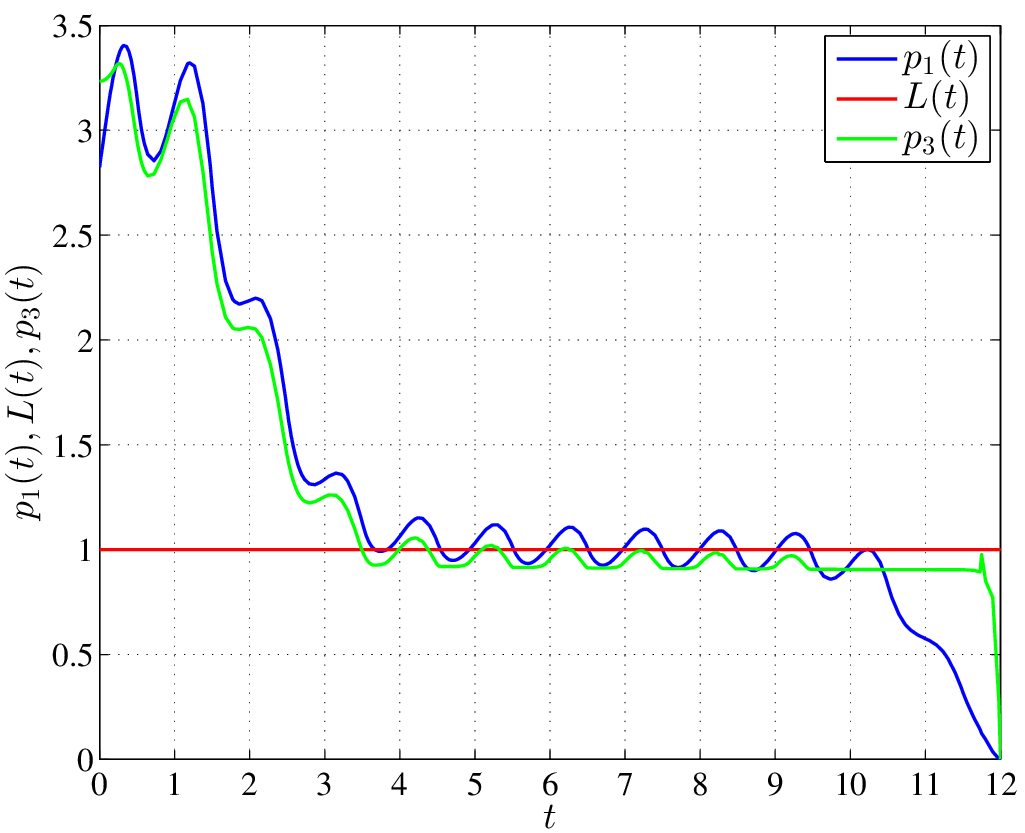}
    \label{fig:MirKozCostate_Annual_Multiple}    
\end{minipage}
\caption{Annual plant with single (left column) or multiple (right column) reproduction periods. A and B are the states, C and D are the corresponding controls and E and F are the costate variables. The maximum photosynthesis rate $f(x)=\frac{1.5x}{1+0.3x}$ and the storage release rate $g(x)= 5 x$, where $x$ is vegetative mass. The rate of destruction of vegetative tissues $\mu(t) = 0.8 \left|\cos\left(\frac{\pi}{12}t\right)\right|$ for the left-hand column and $\mu(t) = 1.8 \left|\cos\left(\pi t\right)\right|$ for the right-hand column. The rate of storage losses $\omega \equiv 0.05$ for the left column; although it was possible to obtain multiple switches for the same $\omega$, this function has been changed in the right-hand column to $\omega(t) = \frac{t}{1+0.5t}$ to better visualise the difference in costates between storage and reproductive output.}
\label{fig:Annuals}
\end{figure*}

We start with the simple scenario of an annual plant with a single reproduction period at the end of the season. The results of the simulation are depicted in the left-hand column of Figure~\ref{fig:Annuals}.

A seed stays dormant for several weeks and then germinates using resources from storage. Germination takes place when environmental conditions are still harsh, to prepare a plant for vegetative growth when these conditions improve. Next, there is a phase of pure vegetative growth, when all assimilated resources are allocated to vegetative mass. 
After an instantaneous switch, all resources are allocated to reproductive mass and vegetative mass decays, reaching a very small value at the end of life equal to 12 months. 
If vegetative mass losses are heavier and cyclic, for example through repeated grazing, after the vegetative growth phase there are multiple switches between vegetative and reproductive allocation, as shown in the right-hand column of Figure~\ref{fig:Annuals}. This is reminiscent of an early model \cite{KiR1982}. In the final stage all resources are allocated to reproductive mass.

\begin{figure*}[p]
\begin{minipage}{0.5\linewidth}
\centering
\caption*{A: Non-forced annual: states}    
\vspace{-3mm}
\includegraphics[scale=0.45]{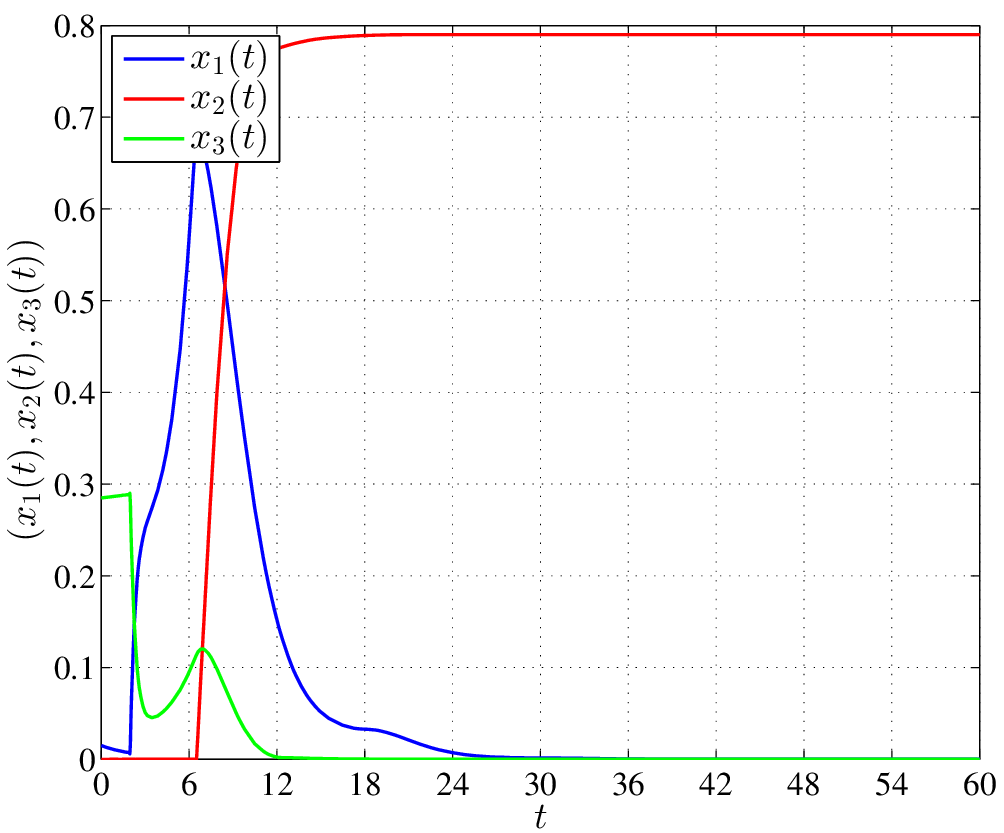}
\end{minipage}
\begin{minipage}{0.5\linewidth}
\centering
\caption*{B: Non-forced biennial: states}
\vspace{-3mm}
\includegraphics[scale=0.45]{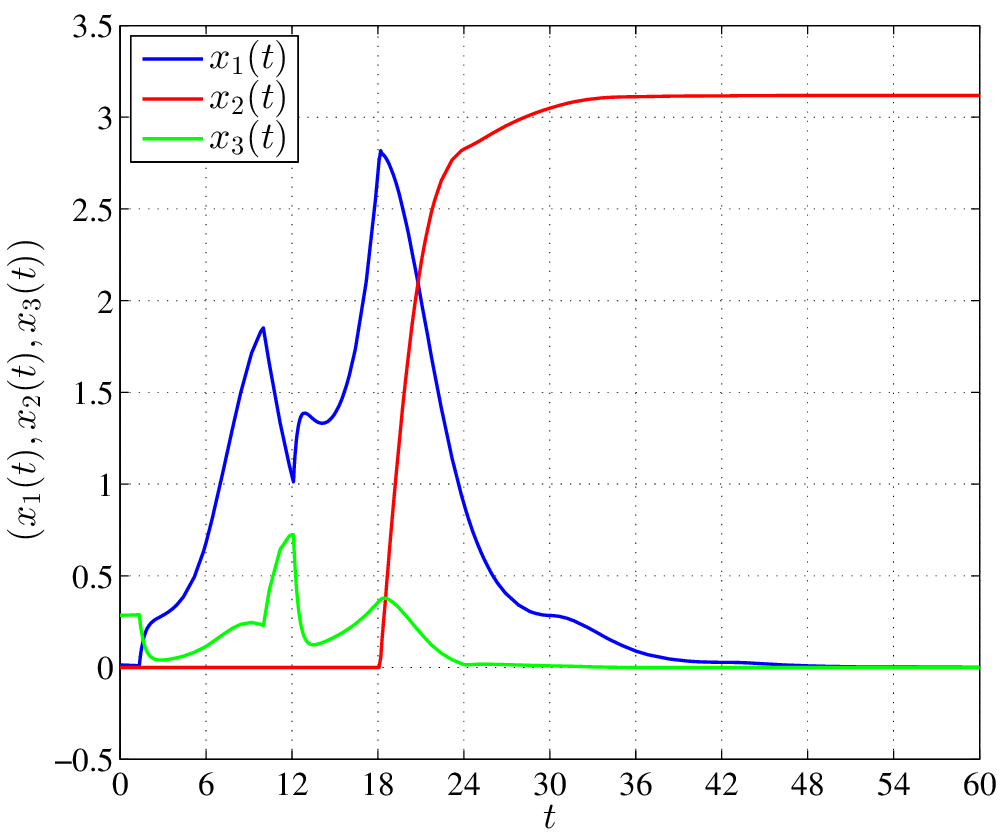}
\end{minipage}

\caption{Optimality of annual (A) and biennial (B) strategies, non-forced by limited lifespan $T$. Storage losses are negligible at the beginning of life and increase very rapidly thereafter. The maximum photosynthesis rate $f(x)=\frac{0.5x}{1+0.01x}$ and the storage release rate $g(x)= 2.5 x$, where $x$ is vegetative mass. The rate of destruction of the vegetative tissues $\mu(t) = 0.8 \left|\cos\left(\frac{\pi}{12}t\right)\right|$ for the annual and $\mu(t) = 0.3 \left|\cos\left(\frac{\pi}{12}t\right)\right|$ for the biennial. The rate of storage losses equals $\omega(t) = 0.000002 t^6$ for the annual and $\omega(t) = 0.0000005 t^5$ for the biennial.}
\label{fig:Nonforced}
\end{figure*}

\begin{figure*}
\begin{minipage}{0.5\linewidth}
\centering
\caption*{A: Monocarp: states}    
\vspace{-3mm}
		\includegraphics[scale=0.45]{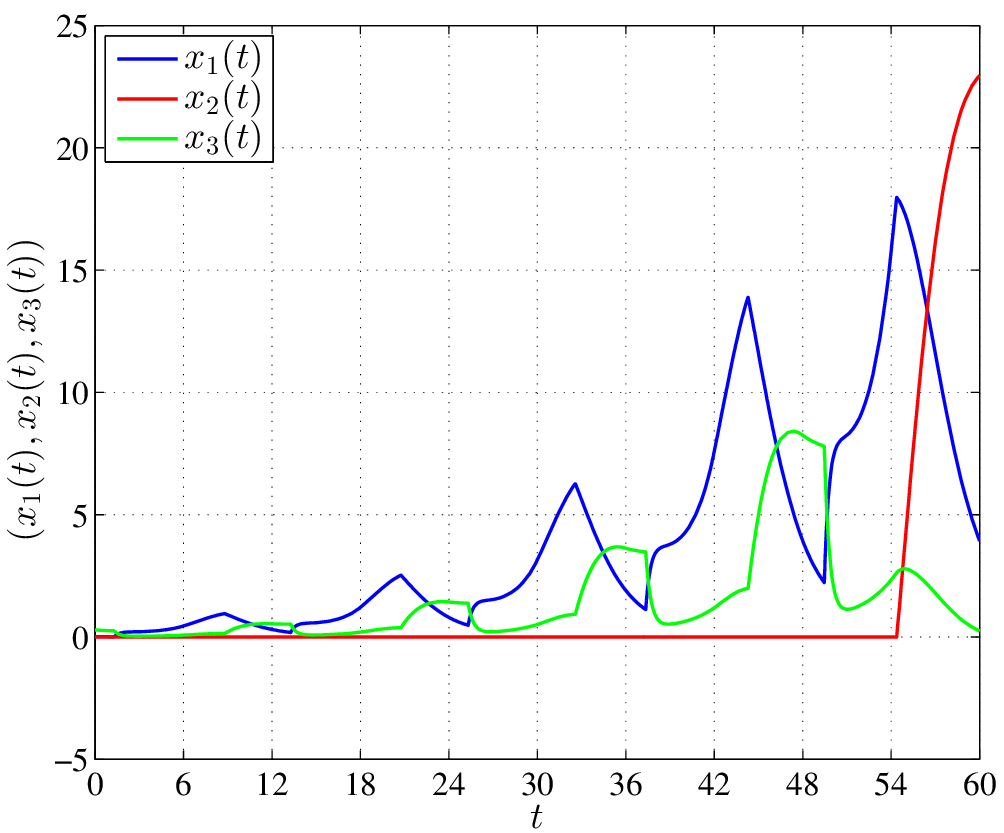}
\label{fig:MirKozState_Mono}
\end{minipage}
\begin{minipage}{0.5\linewidth}
\setlength\parindent{0pt}
\centering
\caption*{B: Monocarp: controls}    
\vspace{-3mm}
\includegraphics[scale=0.45]{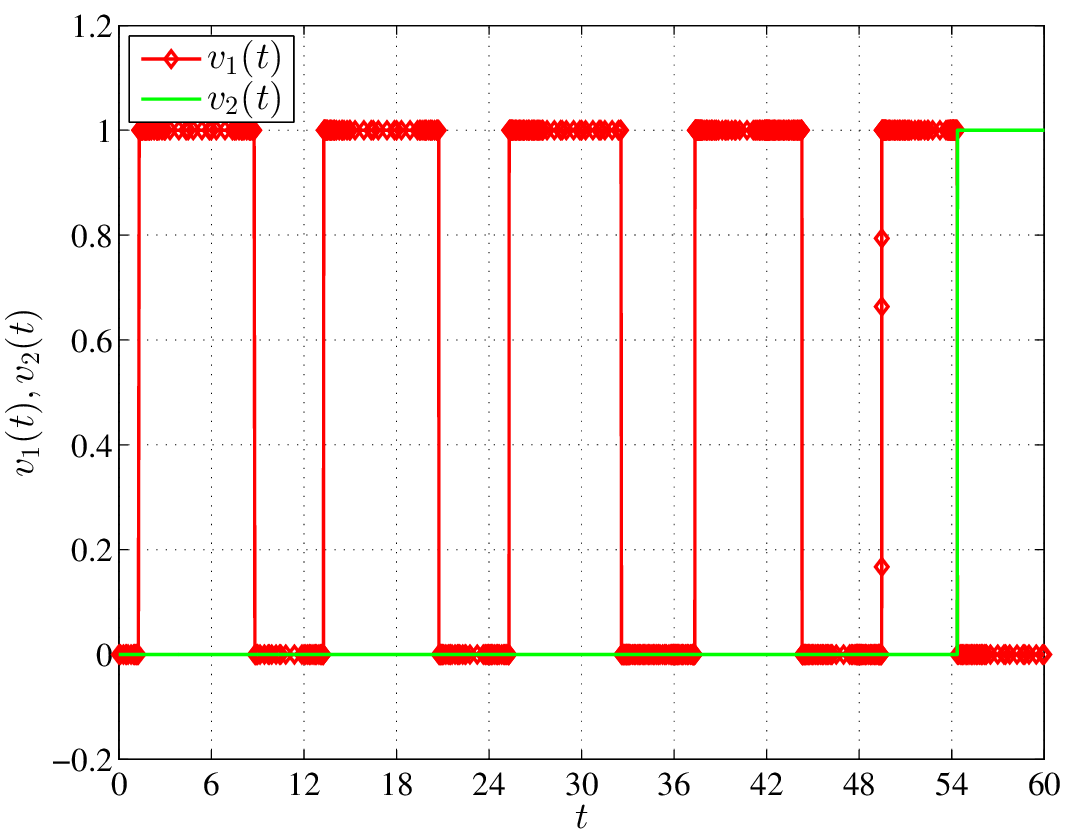}
\label{fig:MirKozControl_Mono}
\end{minipage}
\caption{Monocarp plant with a lifespan set to 60 months. State variables and controls are depicted on A and B, respectively. The maximum photosynthesis rate is $f(x)=\frac{0.5x}{1+0.01x}$ and the storage release rate equals $g(x)= 2.5 x$, where $x$ is vegetative mass. The rate of destruction of the vegetative tissues is $\mu(t) = 0.4 \left|\cos\left(\frac{\pi}{12}t\right)\right|$ and the rate of storage losses $\omega \equiv 0.1$.}
\label{fig:Monocarp}
\end{figure*}

Figure~\ref{fig:Annuals} E and F show the dynamics of costate variables (see Appendix~1) as well as of a survivability function $L$ (which is constant in these examples) for annual plants with one reproductive period and for annuals with multiple reproductive periods. Taking into account that $L$ replaces the costate variable in the case of the reproductive compartment, we can say that resources should be devoted to the compartment with the highest costate variable.

In Figure~\ref{fig:Annuals} E we see that if $p_3$ is not maximal over some time span, then it is only slightly smaller than the maximal costate variable on this interval. To explain this phenomenon, let us assume that at a given time $t$, allocation to reproductive tissues is optimal. If a plant does not behave optimally and does not allocate to reproductive tissues at $t$, resources are retained in storage and can be allocated from storage to reproductive tissues a bit later. Such a small delay decreases fitness minimally, which means that the `usefulness' of allocation to storage is only slightly smaller than the `usefulness' of allocating to vegetative tissues. To enhance the difference between the costate variables for storage and reproduction (Figure~\ref{fig:Annuals} F), we assumed in the annual plant with multiple reproductive periods model that storage losses increase over time.

Life span was set to 12 months for the cases illustrated in Figure~\ref{fig:Annuals}. To show why annual strategy could evolve, we set the lifespan to several years but changed $\omega$, describing storage losses: $\omega$ is now very low during almost the entire season, but it increases rapidly toward the end of the first favourable season. In such a case, the entire amount of storage is relocated to reproductive mass, while only remnants of vegetative mass survive through to the next season and there is no vegetative growth in that season (Figure~\ref{fig:Nonforced} A). 
This is seemingly only virtually annual strategy, but we can easily imagine that a real plant would relocate all movable hydrocarbons to reproductive mass before winter and then die, because remnants of vegetative mass are practically useless. Such relocation, considered by \cite{KoW1986}, was not allowed in the model.

\subsection{Monocarpic plants}
\label{Numerics_Monocarps}

Our next scenario involves a monocarpic plant with a five-year lifecycle. The results are depicted in Figure~\ref{fig:Monocarp}. Since a plant starts its development during winter, it does not grow because the loss of storage is much lower than the loss of vegetative tissues. At the beginning of spring we see the regrowth of the plant from storage, and then vegetative growth due to photosynthesis, which is followed by allocation to storage as preparation for winter. This pattern of development continues until the last season, which ends with reproduction.

An increase in parameter $\omega$ leads to the earlier regrowth of a plant from storage. 
In the limit it leads to the evergreen perennial, as shown in the next subsection.

Life span was set to 60 months for the case illustrated in Figure~\ref{fig:Monocarp}. To show why a monocarpic strategy could evolve, we set the lifespan to several years, and although small at the beginning, $\omega$ increases rapidly with age. Conspicuous inflorescence of monocarpic plants may attract not only pollinators but also enemies, which may cause a rapid increase of $\omega$ with the onset of flowering \cite{JoK05}. Now the first season is used for vegetative growth and then for building storage ({Figure~\ref{fig:Nonforced} B). In the second year, regrowth from storage and then growth from assimilated resources take place, followed by complete allocation to reproductive mass. Although remnants of vegetative mass survive through to the next season, we can call the plant a monocarpic biennial because there is no vegetative growth in the third year.

\subsection{Perennial polycarpic plants}

Figure~\ref{fig:Perennials} presents numerical examples of perennial polycarpic plants. In the case shown in the upper row, plant mortality is neglected, but storage losses are moderate. After a few years of pure vegetative growth, with the peak size of a plant increasing, there are years of mixed vegetative and reproductive growth. Peak size is constant over several years and it decreases close to the end of life, which is arbitrarily set at 96 months. Without such limitation, yearly cycles would be repeated infinitely. Each cycle starts with regrowth from storage during the end phase of an unfavourable season, then the growth of vegetative parts from the current photosynthesis, followed by an instantaneous and complete switch to reproductive allocation (if it appears), followed by the instantaneous and complete switch to building storage. In parallel, vegetative mass decreases during both reproductive allocation and building storage, and this decline continues during an unfavourable season. Only a small amount of vegetative mass persists over an unfavourable season, but stored resources allow for quick regrowth in the next season. Note that the size of storage which is optimal for flowering is reached in the winter before the first reproduction phase.

The middle row in Figure~\ref{fig:Perennials} represents an evergreen plant. Here, storage losses are so high and losses of vegetative mass are low enough so that a plant can survive during an unfavourable season in the form of vegetative parts and storage is not being built. Because the variable $x_3$ represents not only storage but also free sugars still not allocated to another compartment, $x_3$ is not exactly equal to zero. Note that essentially the pattern of growth of an evergreen perennial plant resembles the life history of an annual with multiple reproductive periods, except for longer multi-year life.

\begin{figure*}
\begin{minipage}{0.5\linewidth}
\centering
\vspace{5mm}
\caption*{A: Perennial plant: states}    
\vspace{-3mm}
\includegraphics[scale=0.45]{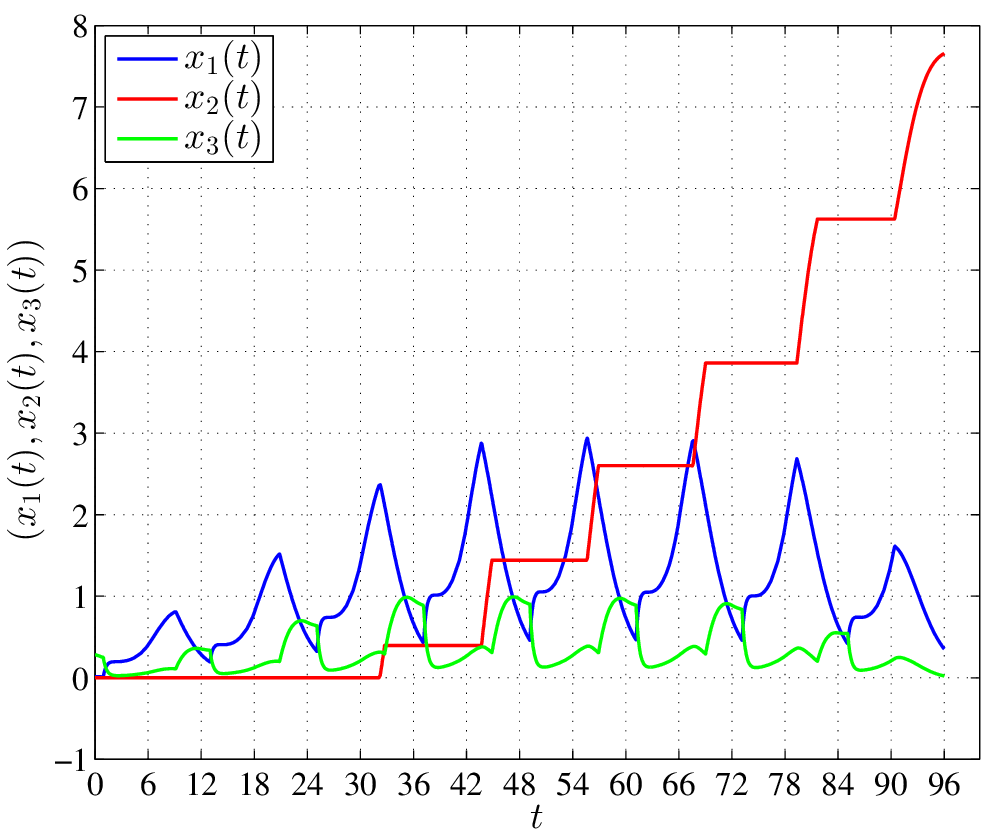}
\end{minipage}
\begin{minipage}{0.5\linewidth}
\setlength\parindent{0pt}
\centering
\vspace{5mm}
\caption*{B: Polycarpic plant: controls}    
\vspace{-3mm}
\includegraphics[scale=0.45]{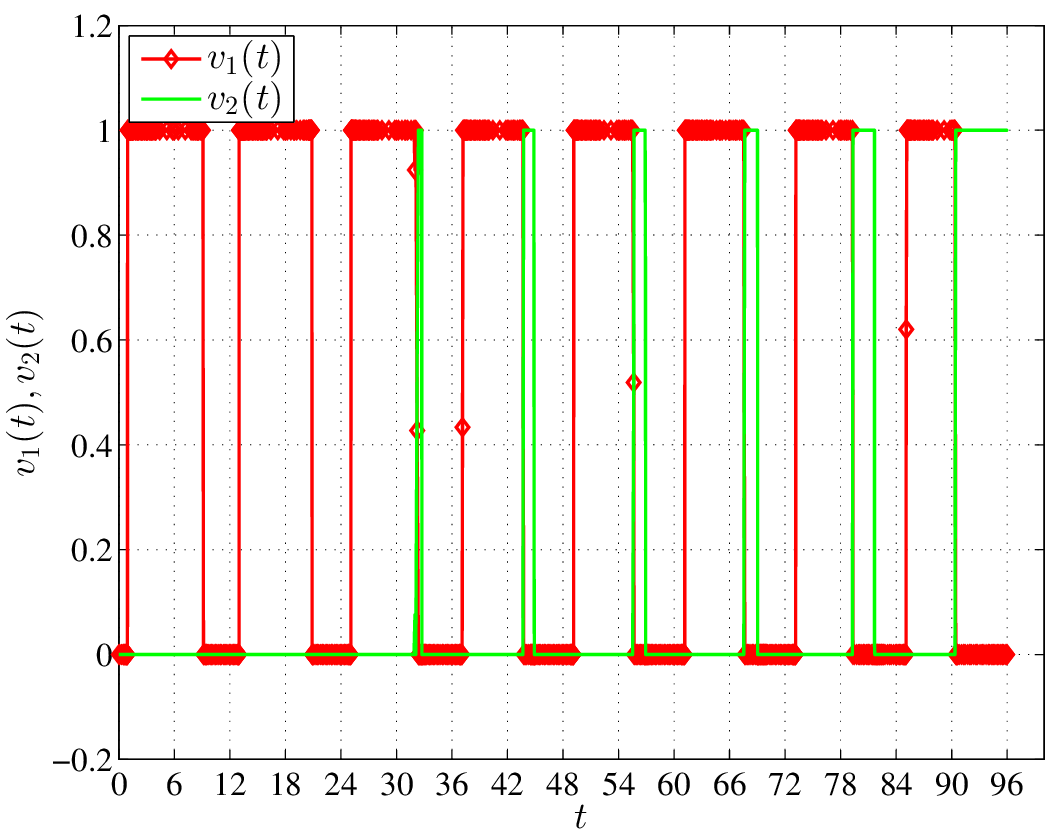}
\end{minipage}

\begin{minipage}{0.5\linewidth}
\centering
\vspace{5mm}
\caption*{C: Evergreen polycarp: states}    
\vspace{-3mm}
		\includegraphics[scale=0.45]{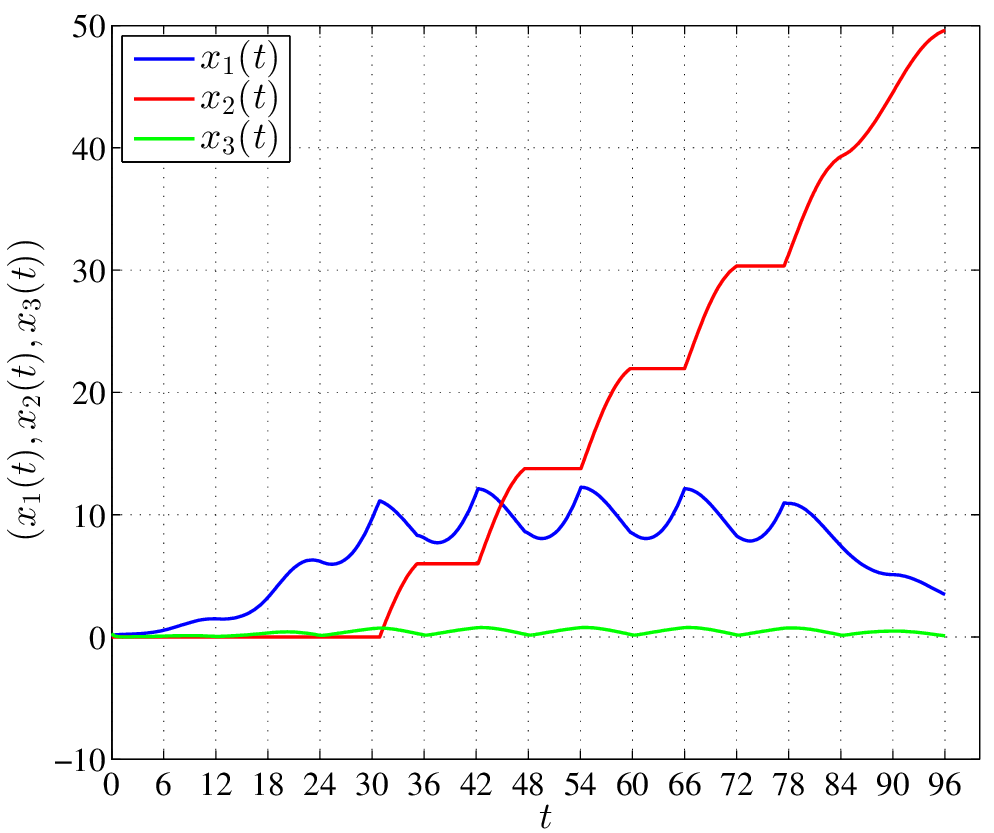}
\end{minipage}
\begin{minipage}{0.5\linewidth}
\setlength\parindent{0pt}
\centering
\vspace{5mm}
\caption*{D: Evergreen polycarp: controls}    
\vspace{-3mm}

\includegraphics[scale=0.45]{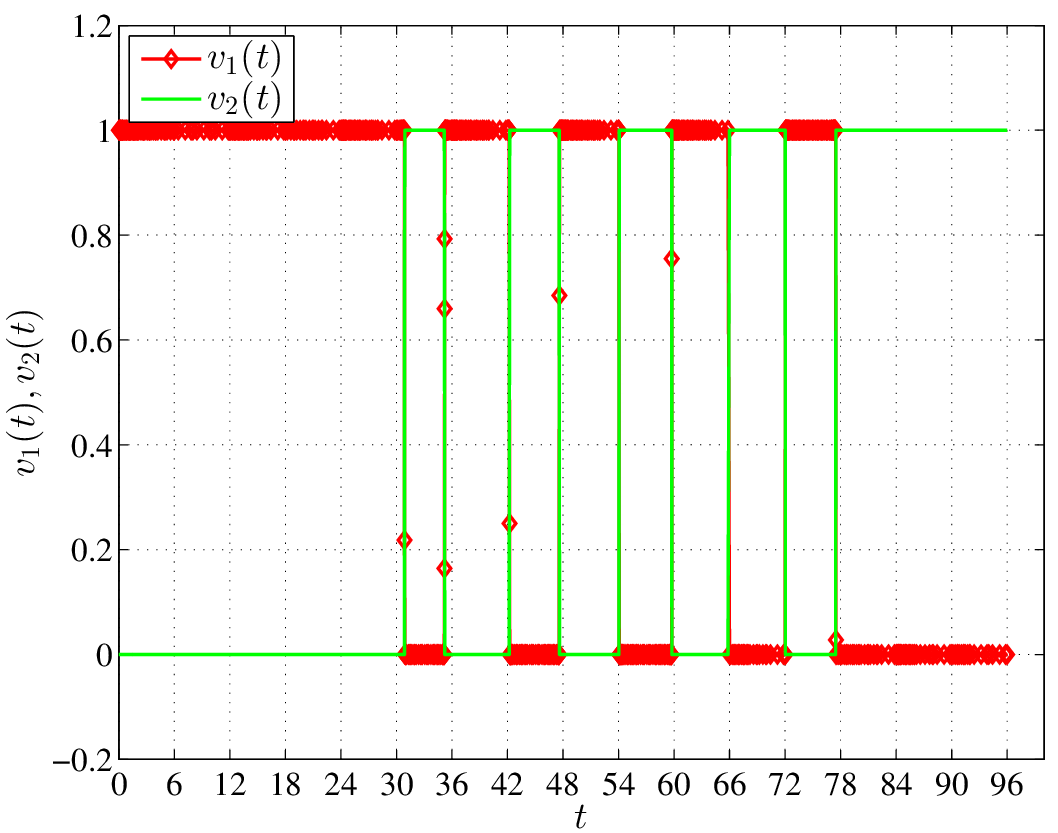}
\end{minipage}

\begin{minipage}{0.5\linewidth}
\centering
\vspace{5mm}
\caption*{E: Polycarp with mortality: states}    
\vspace{-3mm}
		\includegraphics[scale=0.45]{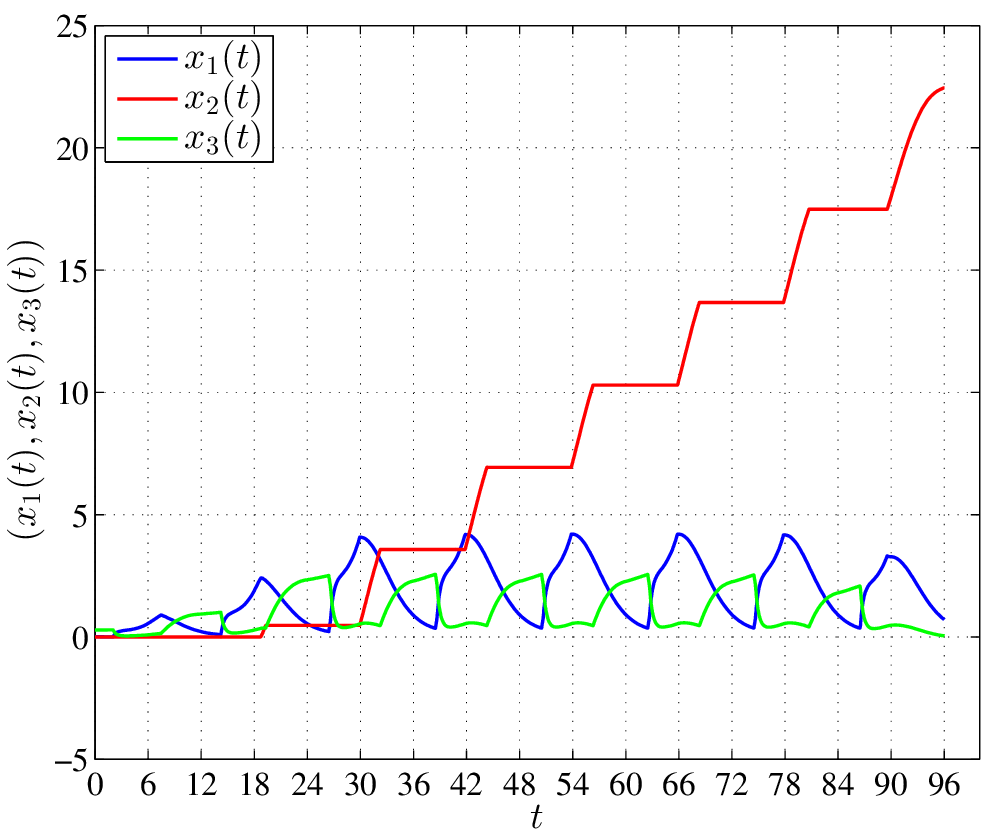}
\end{minipage}
\begin{minipage}{0.5\linewidth}
\setlength\parindent{0pt}
\centering
\vspace{5mm}
\caption*{F: Polycarp with mortality: controls}    
\vspace{-3mm}
\includegraphics[scale=0.45]{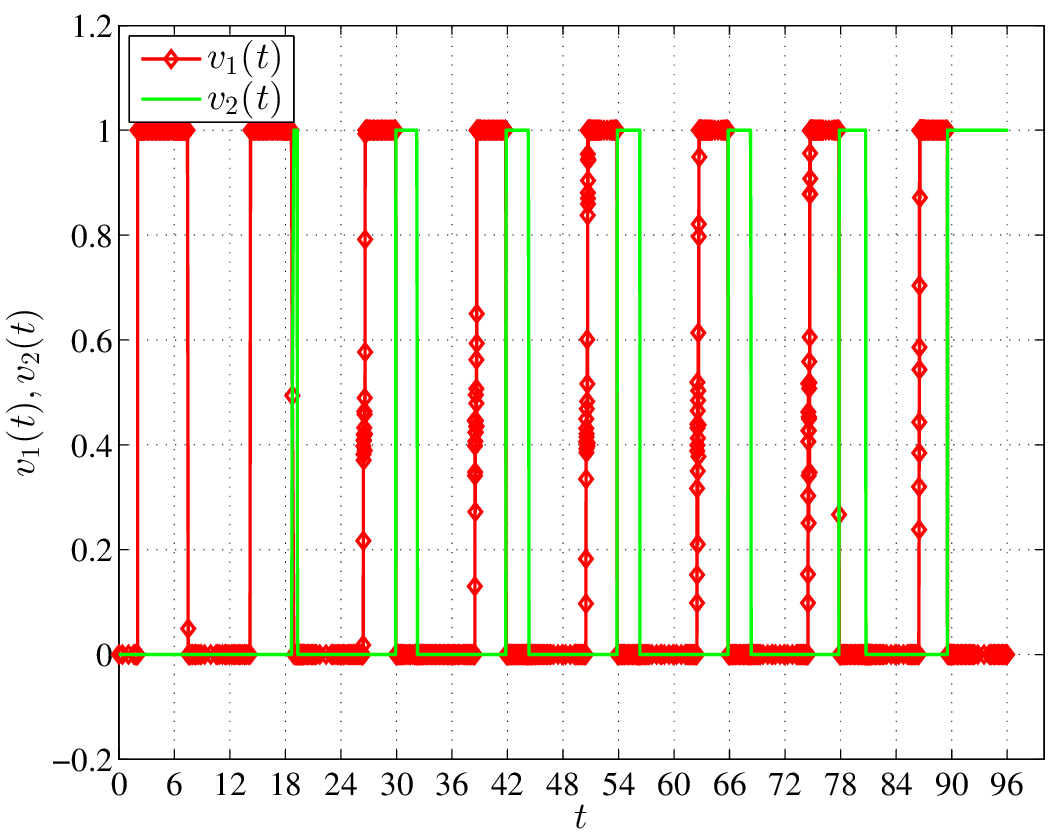}
\end{minipage}
\caption{Perennial plant, with mainly storage persisting over an unfavourable season (A, B, E, F), and an evergreen with permanent vegetative parts and no storage (C and D). In A-D mortality is neglected, but storage losses are moderate for A and B ($\omega \equiv 0.15$) or high for C and D  ($\omega \equiv 1$). In E and F storage losses are neglected, but a constant mortality rate of 0.03 is assumed (probability of survival to age $t$, $L(t) =e^{-0.03 t}$); note that the results for A-B and E-F are qualitatively similar. A, C and E are the states and B, D and F are the corresponding controls. The maximum photosynthesis rate $f(x)=\frac{0.5x}{1+0.1x}$ and the maximum storage release rate $g(x)= 2.5 x$, where $x$ is vegetative mass. The rate of destruction of vegetative tissues $\mu(t)= 0.4 \left|\cos\left(\frac{\pi}{12}t\right)\right|$ for A, B, E and F and $\mu(t) = 0.1 \left|\cos\left(\frac{\pi}{12}t\right)\right|$ for C and D.} 
\label{fig:Perennials}
\end{figure*}

The lowest row in Figure~\ref{fig:Perennials} represents a perennial plant that does not lose storage but is subject to a constant mortality rate (i.e. exponentially decreasing survival probability $L(t)$). To show that mortality has a qualitative effect similar to storage losses, we choose the case with $\omega = 0$. A plant subject to both mortality and storage losses will achieve smaller size, but the general pattern will be the same. Reproductive output is very high in the illustrated case, because it represents plants that survive to the final time $T$. For fitness calculation, $x_2$ is weighted by $L$.

\section{Optimisation of seed mass}
\label{Section_SamenMass}

In the previous sections we defined the fitness of a plant as an expectation of the mass of reproductive tissues produced by the plant during its lifetime. To maximise fitness, a plant controls the allocation of photosynthate and germination time. However, it is well-known that for plants that propagate exclusively through seeds, fitness depends crucially on the quantity and size of the seeds a plant produces. Current models of optimal allocation do not provide this information, and so the mass of a seed is treated as an external parameter.
Essentially, though, choosing the mass of seed is an additional control which a plant can use in order to allocate the photosynthate efficiently. Therefore in this section we 
extend the model from the previous one by giving the plant additional control over the mass of seed.

Let $y_0=(y_1^0,y_2^0,y_3^0)$ be the total mass of seeds (the vector consisting of the masses of three components of a plant) that has to be divided between $a$ seeds, while $a \in [1,\infty)$ and $a$ can be either natural or real number. 

We assume that the mass of each seed is $s=\frac{y_0}{a}$.

The equations determining the dynamics of a plant are as follows:
\begin {equation}
\label{OptSamenMasseModell}
\begin {array} {l}
\dot{x}_1 = v_1(t) g(x_3) - \mu(t)x_1, \\
\dot{x}_2 = (v(t)-v_1(t)) g(x_3),   \\
\dot{x}_3 = \zeta(t)f(x_1) - v(t) g(x_3) - \omega(t)x_3, \\
x(0)=\frac{1}{a}y_0.
\end {array} 
\end {equation}

Here $x(0)=(x_1(0),x_2(0),x_3(0))$.

We look for the values of control variables that maximise the total mass of reproductive tissues produced by all direct descendants:
\begin {equation}
\label{PflZielOptMasse}
\mathop {\max} \limits _ {0 \leq v(t) \leq 1,\ 0 \leq v_1(t) \leq v(t), \ a \in [1,\infty)} 
Q_a = a \xi \int_{t_0}^T L(s) \dot{x}_2(s)ds,
\end {equation}
where a constant $\xi$ is the fraction of germinating seeds. We assume that $\xi$ does not depend on the size of a seed.

In contrast to the problem (\ref{FreiFotModell}), (\ref{Ziel}) with the fixed mass of the seed, the problem (\ref{OptSamenMasseModell}), (\ref{PflZielOptMasse}) may have no solution, in the sense that every finite amount of seeds will not be optimal, which means (with a slight abuse of mathematical rigorousness) that the size of the seeds should be infinitely small ($a \rightarrow \infty$), i.e. no admissible controls $v$, $v_1$ and parameter $a$ generate the optimal value of $Q_a$.

Recall that a function $f$ is called concave on the set $M$, if $\forall z_1,z_2 \in M$, $\forall \alpha \in [0,1]$ an inequality
\begin {equation}
\label{Concave}
f(\alpha z_1 + (1- \alpha)z_2) \geq \alpha f(z_1) + (1-\alpha)f(z_2)
\end {equation}
holds. If the inequality \eqref{Concave} holds with $\leq$ instead of $\geq$, then the function $f$ is called convex on set $M$.

From biological viewpoint important is the case when $f$ and $g$ are concave functions, i.e. the rate of photosynthesis and the maximal speed of chemical reactions in a plant are saturated as a result of the growth of the mass of a plant (due to self-shading of leaves, nutrient depletion in the soil, etc.). For such functions, together with condition $f(0)=g(0)=0$, $a \in [1,\infty)$, one can prove (see Appendix~2) that $Q_a$ increases when $a$ increases.

It follows from this result that when $f$ and $g$ are concave, the best strategy for a plant is to produce as many seeds as possible, which means that the seeds should be as small as possible. A similar argument shows that for convex functions $f$, $g$, the optimal mass of the seed has to be as large as possible (without additional restrictions on the quantity of seeds $a=1$).

The boundary case occurs when both $f$ and $g$ are linear functions, which are concave and convex at the same time. In this case it follows from (\ref{OptSamenMasseModell_Neuform}) that the yield of a plant does not depend on the mass of the seeds.

Another observation is that if $f$ and $g$ are concave and continuously differentiable in the neighbourhood of $0$, then the solution of the linearised model (\ref{OptSamenMasseModell}), (\ref{PflZielOptMasse}) provides the `theoretical' upper bound for the fitness of the plant under consideration.


\section{Discussion}
\label{Besprechung}

Cohen \cite{Coh1971} produced the first explicit model for optimising the allocation of limited resources to growth or reproduction, and this model was applies to annual plants, with the photosynthesis rate dependent linearly on vegetative mass. Denholm \cite{Den75} confirmed Cohen's result through the PMP method. The model was generalised to a non-linear photosynthesis rate by Vincent and Pulliam \cite{ViP1980} and Zi\'{o}{\l}ko and Koz{\l}owski \cite{ZiK83}. Both of these papers used the PMP method to analyse the problem, and in the last one mortality was introduced to the model. King and Roughgarden \cite{KiR1982} also considered vegetative mass losses and  showed that multiple switches from vegetative to reproductive allocation may be optimal if these losses are heavy -- the result that we were able to reproduce in our general model. The first papers on optimal allocation in perennial plants appeared later in \cite{Pug87}, \cite{Pug88} and \cite{IwC1989}. 

Our model combines several specific models within the same framework, which is a great advantage. This goal was achieved through a crucial modification in comparison to the previous models: photosynthates are not allocated directly to vegetative and reproductive tissues, but firstly to storage, from which they could be relocated to other compartments, if optimal at a given time. Such a redefinition of a storage compartment, also including sugars just produced, is biologically reasonable as well as fruitful for mathematical modelling. Analytical analysis of the model leads to the development of general schemes of life history phases, illustrated in Figure~\ref{fig:DiagPlantsBIG}. Such schemes will be very useful in building more advanced models, as well as in planning field studies. The analysis of the model confirmed that optimal switches are instantaneous, in other words resources should be allocated to only one compartment at any given time $t$. 
Resources should go to the vegetative compartment if $p_1$ is larger than $L$ and $p_3$, to storage if $p_3$ is larger than $L$ and $p_1$ and to reproduction if $L$ is the largest of the three.
Simultaneous allocation would be optimal only if two (or more) of functions $p_1,L,p_3$ were equal, which never appears in the presented model. However, we showed in numerical examples that the difference between costate variable for storage $p_3$ and a maximum of $L$ and $p_1$ is fairly small, which indicates that the price for suboptimal strategies can be very low. 
Thus, we should not expect strictly bang-bang switches in nature. Some constraints can also force the optimality of non bang-bang solutions, as discussed later in this paper.

Apart from the most important novelty, i.e treating storage as the primary sink for photosynthates, we added several other novelties to the model. Seasonality was modelled by changing the photosynthesis rate, vegetative tissue losses and storage losses over time. Although we used simple periodic functions in our numerical examples, functions extracted from real data could be applied as well. Such functions could take into account a common in some geographical regions mid-summer depression, i.e. decreasing the rate of photosynthesis in summer months, usually caused by water limitation (e.g. \cite{RoB05}), as well as early decreases in light penetration down to the forest floor, which forces some perennials to bloom and to set seeds early in spring. 
In other papers on optimal allocation, seasonal changes have not been gradual: after a favourable season, the onset of winter is rather abrupt, and then the next favourable season appears instantaneously. 
Such an approach precludes discovering an interesting phenomenon: under some assumed storage losses and vegetative parts, germination or spring regrowth may still appear in an unfavourable season, which allows for using the photosynthetic potential of a favourable part of the season in full. If losses of vegetative parts in winter are low, an evergreen strategy without producing storage is optimal, as shown in Figure~\ref{fig:Perennials} C-D. Such perennials are common, but earlier models were not able to show this strategy because of the non-realistic treatment of seasonality. Note that the transition from the strategy of persisting over winter in the form of almost an exclusive storage organ to an evergreen strategy with virtually no storage is gradual -- if we decrease the losses of vegetative tissue with respect to storage losses, regrowth from storage becomes gradually earlier. Storage losses were not considered earlier in allocation models, except in the context of storage as a backup for an unpredictable and basically a-seasonal environment \cite{IwK1997}.

Although we do not show such a result, annuals starting to grow in autumn and continuing the process through to spring, as in the case of winter cereals, would be optimal under some specific loss functions for vegetative parts.

The interpretation of some of the terms used in this paper could be more general. As stated previously, `storage' ($x_3$ in the model) contains real storage in the form of starch or fat, as well as free carbohydrates just produced. Similarly, losses of vegetative parts may include not only grazing or the decaying of leaves, but also a decrease in photosynthetic potential typical of ageing leaves. 
Storage losses may mean a usage of a part of storage for metabolic processes, but also partial consumption of storage organs by animals.

Our model, as each model, has its limitations. Assuming that $f$ depends only on vegetative mass ($x_1$ in the model), we neglect the possibility of photosynthetic activity in reproductive tissues. Such a possibility was considered by King and Roughgarden \cite{KiR1982}, albeit only for annual plants. Since we suppose that $g$ is a function of $x_3$, the rate of allocation from storage to vegetative and/or reproductive tissues is limited only by the mass of storage. This leads to the rapid regrowth of a plant from storage, because the allocation rate can be very high, even if the mass of existing vegetative tissues is extremely small. To exclude this effect one could consider more general functions $g$, depending also on $x_1,x_2$. Another possibility, investigated in \cite{ZiK1995}, \cite{IoG2005}, is to introduce additional constraints to the allocation rate from one type of tissue to another. However, this issue is outside of the scope of this paper. We also do not allow for the relocation of resources from vegetative mass to reproduction or storage, which could be optimal when the rate of photosynthesis drops to a very low level \cite{KoW1986}.

An important and special case for perennial plants are monocarps. At present, no unified theory explains the origins of monocarpy \cite{KKI97, Iwa00}, but in this paper we provide several possibilities for the emergence of a monocarpic strategy. 

The first sufficient condition for the genesis of a monocarpy, shown in Appendix~1, is the negligibility of mortality and storage losses. In this case the strategy is not forced by the choice of the length of life $T$, and qualitatively it does not depend on the choice of functions $f$ and $g$. 
In this case for a plant there is no sense in reproducing earlier because it can save energy in the storage and, since storage cannot be lost, use it later for producing either vegetative or reproductive tissues. 
One may argue that the mortality of a parental plant usually cannot be neglected for perennial species, but if we also add losses of reproductive tissues, then the negligibility of storage losses may still lead to the development of a monocarpic lifecycle.

Monocarps (or annuals) may also arise from our model when storage losses are moderate, exemplified in Figures \ref{fig:Annuals} and 
\ref{fig:Monocarp}. 
However, these strategies are in a sense forced by limited life span, as the plants have no choice but to reproduce at the end of their life. Such a scenario is believable if certain constraints are not allowing the plant to live longer; for example, if a plant species has no morphological or physiological mechanisms allowing it to survive during the winter, it must be annual. However, we can find many plant genera in which annual, monocarpic and polycarpic species coexist. Even more so, some species represent different strategies in different regions, which means that constraining to one strategy is infrequent. 

A third possible scenario which leads to monocarpy is a rapid increase in storage losses over time. 
As we show in examples illustrated in Figure~\ref{fig:Nonforced}, even without a limited lifespan an annual or monocarpic strategy can evolve if storage losses or mortality rapidly increase. We call these cases non-forced annuals and monocarps. In annuals, producing resistance to frozen storage may have high overhead costs, and necessary resources may be more efficiently used for additional seed production. Monocarps must be equipped with mechanisms allowing for winter survival, so we therefore have to seek another explanation. De Jong and Klinkhamer \cite{JoK05} suggest that mass blooming of monocarps attracts enemies. For example, flowering \textit{Senecio jacobaea} plants have twice as many \textit{Tyria jacobaeae} butterfly egg batches than non-flowering plants. Losses of vegetative and storage organs would be so high that producing additional seeds would increase fitness far more than storing resources for further life. Similarly, \textit{Cygnoglossum officinale} setting seed plants are almost always attacked by the root weevil \textit{Ceutorhynchus cruciger}, whereas non-reproducing plants are almost never attacked. To explain the development of monocarpy for such species, one may consider storage losses, depending on a mass of reproductive tissues, as done in \cite{KKI97}. It is possible to include such behaviour in our model by taking $\omega=\omega(t,x_2)$. We hope that an analysis of this generalisation of our model will provide new insights into the genesis of monocarpy in perennial plants.

We note that the optimal size of a seed depends crucially on the form of the functions $f$ and $g$. For concave functions that are often used to take into account self-shading, the boundedness of resources, etc. (see e.g. \cite{IwC1989}), we have proved according to our model that seeds have to be as small as possible. 

For plants living in open environments and for species occupying early phases in succession (colonising species), the assumption of concavity is not an oversimplification. The behaviour that our model predicts, namely that the optimal strategy is to produce a vast amount of small seeds, is typical for these species \cite{HLM1970}. However, in closed and shady environments, under mineral shortage conditions, or if there is strong competition from established vegetation, the rate of photosynthesis per unit of mass can increase with the increasing mass of the plant, i.e. function $f$ is convex on some $[0,p]$, $p>0$ and the seeds cannot be too small. These predictions are, in general, in accordance with experiments \cite{JaE2000} and \cite{WLL1996}, but see \cite{MGT1998}.
 
Note that a similar result for offspring size in animals was obtained by Taylor and Williams \cite{TaW84} and Koz\l owski \cite{Koz96}: in a-seasonal environments, offspring should be as small as possible if the dependence of production rate/mortality rate is concave, and they should have some optimal size if dependence is convex. Because in their models the necessary condition for the existence of optimal adult size is the concavity of the ratio, optimal size for both adults and offspring can appear only if the ratio has an inflection point(s). 


Another important question is assessing how realistic are fitness measures \eqref{Ziel} and \eqref{PflZielOptMasse}. Such measures of fitness, known also as `lifetime offspring production', are reasonable only in stable populations. In populations changing their size, the timing of reproduction is also important. Descendants which have been produced earlier can in turn produce their own offspring earlier, and thus they are more valuable than descendants produced at a later stage. 
This indicates that the right measure of fitness of the plant is not the production of offspring over a lifetime, but lifetime offspring production by this plant and its descendants. This leads to a considerably more complicated (and more challenging) optimal control model, which to our knowledge has not been addressed in the literature for continuous-time models of plants. Much more frequent is a simpler version of fitness, defined as a number of descendants discounted to the present \cite{MyD95}:
\begin {equation*}
\label{Fitness-measure}
\int_{t_0}^T W(s) (v(s)-v_1(s)) g(x_3(s)) ds  \rightarrow \max,
\end {equation*}
where $W(s)=L(s) e^{-rs}$, for all $s \geq 0$, $r$ is the population growth rate \cite{HoM92}, \cite{KaS93}, \cite{Koz93}. 
If we assume that $r \geq 0$, then $W$ is a decreasing function of time with $W(0)=L(0)$, and an analysis of life-histories -- as performed in this paper -- is still valid for fitness measures \eqref{Fitness-measure} if we change $L$ to $W$.

Despite some limitations, our model, which unifies previous specific allocation models, provides deep insights into a broad range of plant strategies. 
Restrictions on the functions $f$ and $g$ as well as on other functions describing seasonality in \eqref{FreiFotModell} are very mild and allow for finding the right parametrisation of the model for a particular species or group of species. This makes the model helpful for planning experiments and field measurements.

\section{Conclusions and outlook}
\label{Schluss}

In the present paper we have developed a model that describes the optimal allocation strategies of a perennial (as well as an annual) plant during all the stages of its life. The model was analysed with the help of Pontryagin's Maximum Principle, and as a consequence we have derived a description of the life of a perennial plant (Figure \ref{fig:DiagPlantsBIG}) from dormancy until its death. This model encompasses the models of annual as well as monocarpic plants. The sufficient condition for monocarpicity when there is no mortality has also been presented.

In Section~\ref{sec:Numerics} we have shown by means of numerical simulations that the patterns of development of annual plants with one or multiple reproduction periods, monocarpic, evergreen and polycarpic perennial plants, can be obtained through different choices of model parameters. Due to mortality and/or by increasing the destruction rate of storage, one can obtain `non-forced' annuals as well as monocarps.

In Section~\ref{Section_SamenMass} we analysed the trade-off between size and the number of seeds for optimal allocation problems. We have provided sufficient conditions to ensure that an optimal strategy will produce as many (or as few) seeds as possible.

The applicability of our results has been discussed in Section \ref{Besprechung}.

Although our model encompasses a wide variety of different patterns of plant development, it is inapplicable to certain types of plants. In particular, it seems that it is hardly possible to explain the life histories of plants which at the beginning of the season produce flowers and only then start to develop vegetative tissues. It seems that competition for pollinators should be introduced to the model in order to explain such a strategy.

Plants with vegetative reproduction are entirely beyond the scope of this model. In spite of the commonness of such plants in nature, only a few papers are devoted to studying their life strategies \cite{Ole2001}, \cite{Ole2003}. The development of a unifying model for plants with vegetative and sexual reproduction is also a challenging direction for research.

Another interesting topic involves generalising the results in Section~\ref{Section_SamenMass} to the case when the survivability of a seed depends on its size, which will be more realistic from the biological point of view. It seems that methods different from those used in our paper will be required to address this problem accordingly.

We have assumed in the paper that mortality is age-dependent, but size-independent. Introducing size-dependent mortality into the model \eqref{FreiFotModell} makes analysis with the aid of PMP considerably more complex.

Due to a number of results achieved with the help of this model, as well as due to a variety of possible directions for future investigations, we believe that this paper is a good starting point for a fruitful research program in this area.


\section*{Acknowledgments}

We thank Filip Kapustka and Volodymyr Nemertsalov for fruitful discussions and constructive suggestions.

The computations have been made in Matlab, with the help of the optimal control solver GPOPS. JK was funded by Jagiellonian University (DS/WBiNoZ/INoS 757/13).

\section*{Appendix 1: Model analysis}
\label{S:Analysis}

For analysis of \eqref{FreiFotModell} we exploit Pontryagin's Maximum Principle (see, e.g. \cite{ATF87}).
Note that we can drop the equation for $x_2$, since the other equations of \eqref{FreiFotModell} as well as the cost functional \eqref{Ziel} after substitution of $\dot{x}_2$ do not depend on $x_2$.

The Hamiltonian of \eqref{FreiFotModell}, \eqref{Ziel} is defined by:
\begin {equation}
\label{Hamiltonian}
\begin{array}{l}
{H{=} p_1(t) \left(v_1(t) g(x_3(t)) - \mu(t)x_1(t)\right)} \\
   {\phantom{aaaa} + \lambda_0 L(t)\left(v(t)-v_1(t)\right) g(x_3(t))}\\
{\phantom{aaaa}  {+}p_3(t) \left(\zeta(t)f(x_1(t)) {-} v(t) g(x_3(t)) {-} \omega(t) x_3(t)\right).}
\end{array}
\end {equation}

Here $\lambda_0 \geq 0$ and $p_1,p_3$ are so-called adjoint functions. The equations determining their dynamics will be given later.

To simplify the notation, we will frequently write in equations simply $p_1$, $x_2$, etc. instead of $p_1(t)$, $x_2(t)$, if there arises no ambiguity.

We rewrite expression \eqref{Hamiltonian} in a more suitable form
\begin {equation}
\label{Hamiltonian2}
\begin{array}{l}
{H {=} p_3 \zeta(t)f(x_1) {-} p_1 \mu(t) x_1 {+} g(x_3)\left( v_1(t)(p_1{-}\lambda_0L(t)) \right.} \\
{\left.\phantom{aaaaaaaaaa} +v(t)(\lambda_0 L(t)-p_3)\right) - p_3 \omega(t) x_3.}
\end{array}
\end {equation}

Equations for the adjoint function $p$ are as follows
\begin {equation}
\label{P-Gleichungen}
\begin {array} {l}
\dot{p}_1 = p_1\mu(t) -p_3 \zeta(t) \frac{\partial f} {\partial x_1}(x_1), \\
\dot{p}_3 = {-}\frac{\partial g} {\partial x_3}(x_3) (v_1(p_1-\lambda_0L(t)){+}v(\lambda_0L(t)-p_3)) {+} p_3 \omega(t).
\end {array}
\end {equation}

The corresponding boundary conditions are
\begin {equation}
\label{P-RandBedingungen}
{p}_1(T) = {p}_3(T)= 0.
\end {equation}

If $\lambda_0=0$, then from \eqref{P-RandBedingungen} and \eqref{P-Gleichungen} we obtain that $p_i \equiv 0$ on $[t_0,T]$, from which it follows that all the controls are possible. Let $\lambda_0>0$. We can take in this case $\lambda_0=1$.

To obtain the values of $v,v_1$, we solve the problem
\[
H \rightarrow \max,\ 0\leq v \leq 1,\ 0 \leq v_1 \leq v.
\]
It is not hard to check that its solution is given by
\begin{enumerate}
	\item If $ L(t) -p_3(t)>0$, then $v(t)=1$, and
\begin{equation}
\label{Fall1}
 v_1(t) =\left\{
 \begin{array}{cl}
 v(t)  & \textrm{if } p_1(t) - L(t)  >0, \\
 0 & \textrm{if } p_1(t) - L(t)  <0, \\
 \in [0,v ] &  \textrm{if } p_1(t) - L(t)  =0.
 \end{array}\right.
\end{equation}

  \item If $ L(t) -p_3(t)=0$, then
\begin{equation}
\label{Fall2}
 \begin{array}{ll}
 p_1(t)- L(t) >0 & \Rightarrow v(t)=1, v_1(t)=v(t) \\
 p_1(t)- L(t) =0 & \Rightarrow v(t), v_1(t) \textrm{ - EAC} \\
 p_1(t)- L(t) <0 & \Rightarrow v(t) \textrm{ - EAC}, v_1(t)=0
 \end{array}
\end{equation}

 \item If $ L(t) -p_3(t)<0$, then

\begin{itemize}
	\item if $p_1(t)- L(t) \leq 0$ then $v(t)=v_1(t)=0$.
	\item if $p_1(t) -  L(t) >0$ then
\begin{equation}
\label{Fall3}
 \begin{array}{cl}
 p_1(t)-p_3(t) <0 & \Rightarrow v(t)=v_1(t)=0 \\
 p_1(t)-p_3(t) =0 & \Rightarrow v(t) \textrm{ - EAC}, v_1(t){=}v(t) \\
 p_1(t)-p_3(t) >0 & \Rightarrow v(t)=v_1(t)=1
 \end{array}
\end{equation}
\end{itemize}
\end{enumerate}
Here the abbreviation EAC stands for "every admissible control".

We introduce three main periods characterized by different values of controls:
\begin{enumerate}
	\item[(V)] Vegetative period: $p_1(t)>\max\{ L(t),p_3(t)\}$. In this case $v(t)=v_1(t)=1$, that is the vegetative parts are being constructed with the maximal rate.
	
Equations \eqref{P-Gleichungen} in the vegetative period take form
\begin {equation}
\label{P-Gleichungen 1Per}
\begin {array} {l}
\dot{p}_1 = p_1(t)\mu(t) -p_3(t) \zeta(t) \frac{\partial f} {\partial x_1}(x_1(t)), \\
\dot{p}_3 = -\frac{\partial g} {\partial x_3}(x_3(t)) (p_1(t)-p_3(t)) + \omega(t) p_3.
\end {array}
\end {equation}
	
	\item[(R)] Reproductive period: $ L(t)>\max\{p_1(t),p_3(t)\}$. In this case $v(t)=1,\ v_1(t)=0$ and reproductive tissues are being constructed with the maximal rate.
	
Equations \eqref{P-Gleichungen} within this period take the form
\begin {equation}
\label{P-Gleichungen 0Per}
\begin {array} {l}
\dot{p}_1 = p_1(t)\mu(t) -p_3(t) \zeta(t) \frac{\partial f} {\partial x_1}(x_1(t)), \\
\dot{p}_3 = -\frac{\partial g} {\partial x_3}(x_3(t)) (L(t) -p_3(t)) + \omega(t) p_3(t).
\end {array}
\end {equation}
	\item[(S)] Storage period: $p_3(t)>\max\{p_1(t), L(t)\}$. In this case $v(t)= v_1(t)=0$ and all allocated energy goes to storage.		

The corresponding equations \eqref{P-Gleichungen} take the form
\begin {equation}
\label{P-Gleichungen 4Per}
\begin {array} {l}
\dot{p}_1 = p_1(t)\mu(t) -p_3(t) \zeta(t) \frac{\partial f} {\partial x_1}(x_1(t)), \\
\dot{p}_3 = \omega(t) p_3.
\end {array}
\end {equation}
	
\end{enumerate}

We are going to analyze these periods more deeply and find out in what order these periods can arise in a life of a plant.
To this end we investigate equations \eqref{P-Gleichungen} from the end of the life of a plant.

Controls $v$ and $v_1$ maximize the value of $(v_1(p_1-L(t))+v(L(t)-p_3))$, therefore for optimal $v, v_1$ it holds that
\begin {equation}
\label{FolgeAusH}
(v_1(p_1-L(t))+v(L(t)-p_3)) \geq 0.
\end{equation}

Note that in case, when $\omega(t) \equiv 0$ (that is, if storage parts cannot be destructed due to external factors) this inequality and monotonicity of $g$ imply that $p_3$ is an non-increasing function on $[t_0,T]$.

Let us analyze the behavior of Lagrange multipliers $p_i$ and values of controls at the neighborhood of the time $T$.

If the last period would be vegetative, then the equations, governing the dynamics of $p_1,p_3$ would be \eqref{P-Gleichungen 1Per}. Due to wellposedness of \eqref{P-Gleichungen 1Per}, and since the conditions \eqref{P-RandBedingungen} hold, we obtain, that $p_1(t) \equiv 0$ and $p_3(t) \equiv 0$ in the neighborhood of time $T$. Since $L(t)>0$ for all $t<T$, we come to a contradiction with an assumption that the last period is vegetative. Analogously one can show that also the storage period cannot be the last period of a plant's life.
This proves, that \textit{the last period of a plant development is a reproductive period}.

From equations \eqref{FolgeAusH} and \eqref{P-Gleichungen} using monotonicity of $g$ and inequality $\omega \geq 0$ we have that if for some $\tau \in [t_0,T]$ $p_3(\tau)<0$, then $p_3(t)<0$ for all $t \in [\tau,T]$, which contradicts to \eqref{P-RandBedingungen}. Thus, $p_3 \geq 0$ on $[t_0,T]$. Analogously one can prove that $p_1 \geq 0$ on $[t_0,T]$.

Now let us find out, what period can precede to the reproductive period. According to equations \eqref{P-Gleichungen 4Per} and due to $\omega \geq 0$ we see, that $p_3$ cannot decrease during the reproductive period. Since $L$ is a non-increasing function, we see, that starting in a reproductive period ($p_3 > L$) we cannot obtain $p_3 < L$ at the end of this period. This tells us that \textit{before reproduction period the storage period is impossible}.

If the climate conditions (functions $\mu$ and $\zeta$) are such that $p_1(t)=L(t)$ for all $t \in [t_s,t_1]$ for some $t_s <t_1$, then according to \eqref{Fall1} a plant can have the period with mixed control $v_1 \in [0,v]$ for $t \in [t_s,t_1]$. Although this possibility cannot be excluded in general, such mixed controls can arise only due to very specific climate conditions and we do not separate it as a special period of plant life.

If $p_1(t)- L(t)$ is increasing from the left at $t=t_1$, then one can distinguish one more reproductive period $[t_1-s,t_1)$ for some $s>0$. Throughout this paper we follow the agreement to combine all such periods together with stages with mixed controls between these periods into one reproductive period.

Let $p_1(t)- L(t)$ be decreasing. Then for some time interval preceding to the reproductive period we have $p_1(t) > L(t) >p_3(t)$ and therefore on this time interval a plant has a vegetative period. We call it period $V.1$, in contrast to period $V.2$ characterized by relation $p_1(t) >p_3(t) > L(t)$ (this distinction will be useful for monocarpic plants).

There are 2 possibilities for the plant behavior before period $V.1$: either it will have one more $R$-period (if $p_1$ decreases lower than $L(t)$ while it remains true that $p_3<L(t)$), or it will exist $t_2<t_1$: $p_3(t_2)=L(t_2)$. As mentioned before, we neglect the possibility of mixed controls and consider the case $\dot{p}_3(t_2)<0$.

In this case period $V.2$ characterized by $p_1(t)>p_3(t)>  L(t)$ precedes the period $V.1$.
 Although the allocation pattern is the same in both periods $V.1$ and $V.2$, the distinction between these periods is useful for the study of phenology of monocarpic plants. To understand this difference let us consider the case, when the nonstructural carbohydrates cannot be deconstructed due to external factors (i.e. $\omega \equiv 0$, which implies, as was mentioned earlier, that $p_3$ is non-increasing) and the probability of survival remains constant throughout the whole period ($L \equiv const$). This implies that before period $V.2$ the reproduction periods are not possible ($p_3>L$) and consequently \textit{the plant exploits monocarpic strategy}.

In the general case, when $\omega \not \equiv 0$ both periods $R$ and $S$ can precede the $V$-period, or all the previous life of a plant can consist of one vegetative period. In the first case a plant possesses one more reproduction period, which has been already analyzed. If before vegetative period there is no other period, then the \textit{plant is annual}.

Let now the $S$-period precedes to the $V$-period.
Then there exist $t_4,\ t_3$: $t_4<t_3<t_2$, such that $p_1$ increases on $[t_4,t_3]$ (due to the unfavorable climate conditions) and $p_1(t_4)=p_3(t_4)$.
We separate period between $t_4$ and $t_3$ in the season $V.2.1$ ($p_1>p_3>L(t)$, but $p_1$ increasing), which distinctive feature is that \textit{although the climate conditions are not comfortable for photosynthesis a plant anyway allocates some part of stored resources to the construction of the vegetative tissues, so as to come into the better conditions with a certain amount of already developed vegetative mass}.

Now let there exist some $r$: $p_1(t)<p_3(t)$ for all $t \in [r,t_4)$. Then a plant enters a storage period.

If the climate conditions are unfavorable for all $t<t_4$, that is, $p_1(t)<p_3(t)$ for all $t \in [0,t_4)$, then the first period of time is only the storage of allocated photosynthate (this is hardly possible because a seed has a possibility to stay this period in dormancy). If it is not the case, then there exist some moments $t_6,\ t_5$, $t_6<t_5<t_4$, such that $p_1$ is decreasing on $[t_6,t_5]$ and $p_1(t_6)=p_3(t_6)$.

We separate the period $(t_5,t_4)$, which we call period $S.1$ (when the climate conditions are disadvantageous and all the allocated material is stored), and time-span $(t_6,t_5)$ called period S.2 (when the climate conditions are kindly, but all the allocated material is anyway stored for a preparation to the unfavorable climate conditions).

Both reproductive and vegetative periods can precede to the storage period. It depends on the climate conditions and values of $x^0$.

It seems that in general we cannot say more about the time of propagation and type of the first period. The reason is that one can choose the values of the initial parameters that are biologically inadequate and consequently obtain unrealistic predictions.
For example, if the climate conditions are chosen to be unfavorable for photosynthesis throughout all the time-interval $[0,T]$, then the model is inapplicable, because the strategy to stay in dormancy all the period is not allowed in the model.

To exclude such biologically irrelevant behavior, we considered in the paper only the case, when the first period after sprouting is vegetative.

\section*{Appendix 2: Proof of the main proposition from Section~\ref{Section_SamenMass}}
\label{S:Proof_of_Prop}


\begin{proof}
The problem (\ref{OptSamenMasseModell}), (\ref{PflZielOptMasse}) can be written in equivalent form, using new variables $y_i(t) :=a x_i(t)$, $i=1,2,3$. Then we have:
\begin {equation}
\label{OptSamenMasseModell_Neuform}
\begin {array} {l}
\dot{y}_1 = v_1(t) a g(\frac{y_3}{a}) - \mu(t)y_1, \\
\dot{y}_2 = (v(t)-v_1(t)) a g(\frac{y_3}{a}),   \\
\dot{y}_3 = \zeta(t)a f(\frac{y_1}{a}) - v(t) a g(\frac{y_3}{a}) - \omega(t)y_3, \\
y(0)=y_0.
\end {array}
\end {equation}

The corresponding maximum problem is:
\begin {equation}
\label{PflZielOptMasse_Neuform}
\mathop {\max} \limits _ {0 \leq v(t) \leq 1,\ 0 \leq v_1(t) \leq v(t),\ a \in [1,\infty)}
Q_a = \xi \int_{t_0}^T L(s) \dot{y}_2(s)ds.
\end {equation}

Now the problem is similar to \eqref{FreiFotModell}, \eqref{Ziel}, but with $af(\frac{y_1}{a})$ and $ag(\frac{y_3}{a})$ instead of $f(x_1)$ and $g(x_3)$.

Using concavity we have: $f(\frac{y_1}{a}) = f(\frac{1}{a}y_1+\frac{a-1}{a} \cdot 0) \geq \frac{1}{a}f(y_1)+\frac{a-1}{a} f(0)= \frac{1}{a}f(y_1)$.

Thus, for every $y_1(t) \geq 0$, $a \geq 1$ it holds $af(\frac{y_1(t)}{a}) \geq f(y_1(t))$ and therefore $af(\frac{y_1(t)}{a})$ and $ag(\frac{y_3(t)}{a})$ are non-decreasing in $a$ and $\sup_{a \in [1,\infty)}af(\frac{y_1(t)}{a})$ and $\sup_{a \in [1,\infty)}ag(\frac{y_3(t)}{a})$ yields, when $a \rightarrow \infty$.

Define the optimal trajectories of the problem (\ref{OptSamenMasseModell}), (\ref{PflZielOptMasse}) for a fixed $a$ as $y(\cdot)$.
Now take arbitrary $n>a$ and consider a system
\begin {equation}
\label{OptSamenMasseModell_1}
\begin {array} {l}
\dot{y}_1 = v_1(t) a g(\frac{y_3}{a}) - \mu(t)y_1, \\
\dot{y}_2 = (v(t)-v_1(t)) a g(\frac{y_3}{a}),   \\
\dot{y}_3 = \zeta(t)n f(\frac{y_1}{n}) - v(t) a g(\frac{y_3}{a}) - \omega(t)y_3, \\
y(0)=y_0.
\end {array}
\end {equation}

The solution of this system at time $t$ subject to optimality condition \eqref{OptSamenMasseModell_Neuform} we denote $\hat{y}(t)$. If $\zeta(0)>0$, then from $nf(\frac{y_1(t)}{n})>af(\frac{y_1(t)}{a})$ we have that $\dot{\hat{y}}_3(0)>\dot{y}_3(0,a)$ and therefore there exists $t^*>0:\ \dot{\hat{y}}_3(t)>\dot{y}_3(t,a)\ \forall t \in [0,t^*)$.
Hence $\hat{y}_3(t)>y_3(t,a)$ and $ag(\frac{\hat{y}_3(t)}{a})>a g(\frac{y_3(t,a)}{a})$ for $t \in (0,t^*)$.

Let $v$ and $v_1$ be the optimal controls for the system \eqref{OptSamenMasseModell_Neuform}. There exist controls $0 \leq \hat{v} \leq v$, $0 \leq \hat{v}_1 \leq v_1$ for the system \eqref{OptSamenMasseModell_1}, such that $\hat{v}(t)a g(\frac{\hat{y}_3(t)}{a})=v(t)a g(\frac{y_3(t,a)}{a})$ and $\hat{v}_1(t)a g(\frac{\hat{y}_3(t)}{a})=v_1(t)a g(\frac{y_3(t,a)}{a})$.

Consequently, $\hat{y}_i(t)=y_i(t,a)$, $t \in [0,t^*)$, $i=1,2$. Constructing $\hat{v},\ \hat{v}_1$ for all $t \in [0,T]$, we obtain that $\hat{y}_2(T)=y_2(t,a)$ and thus for a given $a$ and $n>a$ the optimal trajectory of a system \eqref{OptSamenMasseModell_1} produces no less output than the best trajectory of \eqref{OptSamenMasseModell_Neuform}.

Analogously, the output of the following system is not less than that of the system \eqref{OptSamenMasseModell_1}:
\begin {equation}
\begin {array} {l}
\dot{y}_1 = v_1(t) n g(\frac{y_3}{n}) - \mu(t)y_1, \\
\dot{y}_2 = (v(t)-v_1(t)) n g(\frac{y_3}{n}),   \\
\dot{y}_3 = \zeta(t)n f(\frac{y_1}{n}) - v(t) n g(\frac{y_3}{n}) - \omega(t) y_3, \\
y(0)=y_0.
\end {array}
\end {equation}

Hence $Q_a$ is non-decreasing in $a$.
\end{proof}



\bibliographystyle{elsarticle-num}

\bibliographystyle{plain}
\bibliographystyle{elsarticle-harv}

\end{document}